\documentclass[11pt]{article}
\pdfoutput=1
\usepackage[utf8]{inputenc}
\usepackage[T1]{fontenc}
\usepackage{lmodern}
\usepackage[margin=1in]{geometry}
\usepackage{graphicx}
\graphicspath{{./}}
\usepackage{booktabs}
\usepackage{array}
\usepackage{amsmath,amssymb}
\usepackage{microtype}
\usepackage[hidelinks]{hyperref}
\usepackage{xurl}
\usepackage{pdflscape}
\usepackage{xcolor}
\usepackage{listings}
\usepackage{caption}
\usepackage{titlesec}
\titleformat{\section}{\large\bfseries}{\thesection}{0.6em}{}
\titleformat{\subsection}{\normalsize\bfseries}{\thesubsection}{0.6em}{}

\title{\textbf{RADAR: Readiness for AI Discovery and Agentic Reach}\\[4pt]
\large Measuring the Accessibility of Digital Government Services to AI Systems}

\author{Luke Jordan \quad Tiago C. Peixoto \quad Manuel Ramos-Maqueda\\[2pt]
{\normalsize World Bank}\\[10pt]
{\normalsize\itshape Contributors}\\[4pt]
{\normalsize Luukas Ilves, Manuel Kilian, and Simone Maria Parazzoli (The Agentic State);}\\
{\normalsize Tapan Parikh (Cornell Tech);}\\
{\normalsize Ronaldo Lemos, Fabro Steibel, and Celina Bottino (ITS Rio);}\\
{\normalsize Aleksandar Linc-Djordjevic and Marija Novakovic (DSC);}\\
{\normalsize Saniya Ansar, Asami Okahashi, Flavio Sotelo, Rafael Morado, and Christopher Tullis (World Bank)}}
\date{Preprint, July 2026}

\begin{document}
\maketitle

{\let\thefootnote\relax\footnotetext{We thank Dan Rogger and the World Bank's Data for
Better Governance team for organizing discussions on this work that informed its later
iterations. The findings, interpretations, and conclusions expressed here are those of the
authors and do not necessarily represent the views of the World Bank, its Executive
Directors, or the governments they represent.}}

\vspace{-2.5em}

\begin{abstract}
\noindent Governments increasingly meet citizens through an AI system rather than a website. RADAR (Readiness for AI Discovery and Agentic Reach) measures whether that system works, across 166 countries and on two tasks: whether a chatbot can give a correct, officially sourced, country-specific answer about a public service (informational legibility), and whether an automated agent can reach the service to act on it (agent operability). The central finding is that AI can describe public services far better than it can reach them. In every one of the 166 countries, informational legibility scores exceed average agent operability scores under RADAR's respective measures, and the gap does not shrink with national wealth. Income and language explain only part of the pattern, and several governments perform far better or worse than their resources predict. The two failures have different correlates and different fixes. Whether AI can describe a service is associated with how well a country's main administrative language is represented in web-scale corpora, which a government cannot change quickly. Whether an agent can reach it is associated with the country's national web presence, which a government can change now. Traditional digital-government rankings miss the second problem entirely. RADAR lets governments at any income level see it and offers a concrete agenda to fix it, so that public services are not only described by AI but actually reachable through it.
\end{abstract}

\section{Background: Public services and the new intermediary layer}
Not all citizens access digital government services in isolation. Often, and particularly in Lower Middle Income Countries (LMICs), they rely on intermediaries: a cybercafé employee who has been through the process, an accountant who navigates regulations, and other informal brokers who become essential nodes to access digital services (Hoefsloot et al.~2025). These intermediaries translate between what the government offers and what people need to know. They absorb complexity, correct misunderstandings, and route people to the right place.

Now, AI systems are becoming a new intermediary layer between citizens and the world they interact with. Research on ChatGPT usage shows that seeking information and practical guidance represents the dominant use case for many public facing LLM applications, more common than writing assistance (Chatterji et al.~2025).

Government information is no exception. Google's ATLAS study maps 15 million de-identified interactions across Gemini and Google AI Mode in 150 countries. In the United States, the one country where its data can be benchmarked against national time-use statistics, government services and civic obligations are the most disproportionate category of all AI use, appearing at close to twenty times their share of how Americans spend their time, and around half of these conversations happen outside standard working hours (Iscenko et al. 2026). In Australia, 21\% of citizens report using generative AI to find information about government services, rising to 29\% among daily AI users, suggesting the practical ``front door'' to government is already shifting from portals to AI interfaces (Publicis Sapient 2025). In the United States, Pew Research Center finds that when search results include an AI-generated summary, users are half as likely to click through to source websites (8\% versus 15\%), and almost never click the citations within the summary itself (1\% of visits). Government content is disproportionately subject to this mediation: .gov sites account for 6\% of sources cited in AI summaries versus only 2\% in standard search results. Oftentimes, the information reaches citizens through AI, without direct connection to its source (Chapekis and Lieb 2025). This means that a government's content is increasingly consumed through AI mediation, not direct access, and points to an underappreciated policy concern: the legibility of government content to AI systems.

If chatbots and AI summaries are the new informational intermediary, agents are the next operational intermediary. Chat-based systems mediate answers, but agentic systems increasingly mediate users' actions: they search websites, follow links, interact with web interfaces, and attempt to complete tasks on a user's behalf (Deng et al.~2023). This transition shifts the constraint from informational legibility to operational accessibility. What makes a service operable by agents is different from what makes it usable by humans: in an agent-mediated world, design choices that are often invisible to most human users, such as deep links, stable URLs, cross-domain flows, client-side rendering, and bot controls, can determine whether a digital service can be reached at all (Zhou et al.~2024).

This is not a future scenario. Major technology companies are already building the foundations for a new kind of internet, where AI agents can interact directly with websites and online services. Instead of relying on people to click through pages, these agents use standardized tools and conversational interfaces to access information and complete tasks automatically. As a result, commercial websites are increasingly being redesigned so that machines, not just humans, can navigate them efficiently (The Economist 2025). Government digital services exist in this environment whether they adapt to it or not.

RADAR sits within a broader line of work on the `agentic state': the idea that AI agents may increasingly mediate not only access to information about government, but the delivery and execution of public-service tasks themselves (Ilves et al. 2025). This prospect creates a prior measurement question: before governments redesign institutions and service channels for agentic interaction, can today's digital state already be read and reached by AI systems?

Some governments are already moving from informational AI to operational AI. For instance, Ukraine's First Vice Prime Minister and Minister of Digital Transformation, Mykhailo Fedorov, describes the direction as a model where ``just one request or a single voice message stands between a person's need and the result,'' pointing to Diia.AI, an agent-powered services platform, as an early national step toward that approach.

But most governments are likely to lag behind Ukraine, and even those moving toward agentic services have limited visibility into how ready their online presence is. The challenge resembles an earlier transition: a decade ago, many government portals were technically online but effectively unusable on mobile devices. Governments that optimized early were better positioned when mobile became the dominant access channel. Today, a similar transition is underway, but there is no equivalent of mobile-responsiveness testing for AI readiness. A portal can score well on conventional e-government indices while being largely illegible to chat-based AI systems or inaccessible to automated agents. This is the measurement gap that the Readiness for AI Discovery and Agentic Reach (RADAR) index addresses: it assesses whether government services are accessible not only to people, but also to the AI systems that increasingly mediate how people discover information and access services.

RADAR is not the first attempt to measure what AI systems tell citizens about government services, but it is the first to do so across many countries and to pair informational legibility with agent operability. Recent benchmarks cover a single jurisdiction and the chat channel only, for example audits of the institutional defaults that multilingual models assume (Wang and Suresh 2026) and country-specific government-information assistants such as the CitizenQuery benchmark (Majithia et al. 2026). None spans 166 countries, and none tests whether an automated agent can reach the service, which is the binding constraint this paper documents. On the agent side, general web and computer-use benchmarks such as WebArena (Zhou et al. 2024) and OSWorld2.0 (Yuan et al. 2026), where the best agent completes only about a fifth of long-horizon tasks, show how far real-world navigation still falls short, but they do not target government portals or measure them at country scale.

\section{Methodology}
When a citizen turns to an AI system for help with a government service, two things must go right. First, the system must provide usable guidance: specific to the service and jurisdiction, and linking to or anchored in an authoritative source. Second, if the system (or an agent acting on the user's behalf) tries to reach the official entry point, it must be able to navigate there. These are distinct capabilities. A service can be well documented and that documentation well reflected in LLM output, but operationally hard for an agent to reach. A portal can be easy to traverse but contain sparse or outdated guidance that forces the model to guess and not be reflected in model answers. We call the first capability \emph{informational legibility} and the second \emph{agent operability}.

\begin{center}\begin{minipage}{\linewidth}\centering\small
Table 1. The three RADAR evaluation modes.\\[5pt]
\setlength{\tabcolsep}{4pt}\renewcommand{\arraystretch}{1.15}
\begin{tabular}{>{\raggedright\arraybackslash}p{0.155\linewidth}>{\raggedright\arraybackslash}p{0.258\linewidth}>{\raggedright\arraybackslash}p{0.258\linewidth}>{\raggedright\arraybackslash}p{0.258\linewidth}}\toprule
 & Chat & Agent DOM & Agent Visual \\
\midrule
Measures & Informational legibility & Agent operability & Agent operability \\
Task & 20 services, three-turn citizen-style conversations, in English & One service (birth certificate, with passport fallback), localized-language search & Same task as DOM \\
Interface & API chat interfaces & Browser automation reading page structure (code) & Browser automation operating rendered pages (screenshots) \\
Models & GPT-5, Claude, Gemini, DeepSeek & GPT-5, Claude, Gemini, DeepSeek & GPT-5, Claude, Gemini \\
Scoring & Two LLM judges scoring four dimensions (verifiability, specificity, depth, transparency), 0--10 & Six-dimension rubric with evidence requirements, 0--10 & Same rubric as DOM \\
Share of Overall & One third (mean of four models) & One third & One third \\
\bottomrule\end{tabular}
\end{minipage}\end{center}
\subsection{Informational legibility: chat mode}
Informational legibility refers to whether a chat-based AI system can provide correct, specific guidance about a public service, grounded in official sources rather than improvised from model priors or routed through unofficial intermediaries, and with the appropriate level of certainty. We use ``legibility'' to mean whether official guidance is readable and usable by AI systems. Citizens most often encounter this when asking a question of an LLM through a chatbot interface.

In this paper we measure that legibility directly, through what we call chat mode. \emph{Chat mode} presents a standardized conversation about 20 government services involving a query and two to three follow ups to multiple large language models and evaluates each response against the four legibility dimensions (verifiability, specificity, depth, transparency). Scores are averaged across government services and models to reduce dependence on any single service or system's idiosyncrasies. This measures what AI systems collectively ``know'' or can retrieve about government services, independent of whether an agent could navigate to the source.

Chat mode used each model's API chat interfaces to conduct the conversation. The prompts were constructed by the research team to mimic the questions that an ordinary resident would ask a chatbot when seeking information on services, in a three step conversation. The first prompt states a need in plain language. Examples included, ``How do I renew my ID?'', ``I think it's tax season. Where do I submit my taxes?'', and ``I'm worried about cancer. How do I get a test?''. The second prompt asks about online availability, typically in the form, ``Is there a site to find out more?''. The third prompt probes a specific detail, like ``How long does it take?''.

This analysis covers 166 countries spanning low-, middle-, and high-income economies. Data collection took place twice, a first run in December 2025 and a second run in March 2026 after improvements in the methodology. We tested four large language models: OpenAI GPT-5 (via the OpenAI API), Claude (Anthropic), Gemini (Google), and DeepSeek. The results present only the final run. Gemini was run with its ``search mode'' parameter set to true. The other three models do not make that parameter explicit in their APIs, as in their modernized form the models are understood to make the decision on tool use such as search autonomously. For cost reasons, max output tokens were kept limited unless and until there was evidence of truncation, which did occur with Gemini and ChatGPT. Table 2 lists the chat model versions; the DOM and Visual agent runs used the provider versions current in the collection window, documented in the replication repository.

\textbf{Table 2. Models used, Chat Mode.}

\begin{center}\begin{minipage}{\linewidth}\centering\small
\begin{tabular}{llll}\toprule
Model & Provider & Version & Max output tokens \\
\midrule
GPT-5 & OpenAI & gpt-5-2025-08-07 & 8,192 \\
Claude & Anthropic & claude-sonnet-4-6 & 4,096 \\
Gemini & Google & gemini-3-flash-preview & 8,192 \\
DeepSeek & DeepSeek & deepseek-chat & 8,192 \\
\bottomrule\end{tabular}
\end{minipage}\end{center}
The conversations were conducted across 20 government service scenarios organized into seven domains: public administration (ID renewal, driver's license renewal, booking a driver's test, marriage certificate), health (booking a GP appointment, medication restocking, diagnostic testing), employment (job search assistance, government-supported training, unemployment benefits, maternity benefits), education (school enrollment, school results), taxation (filing income taxes, contesting a tax assessment, tracking a refund, claiming deductions), public safety (reporting a crime, checking case status), and general services (locating the national e-services portal).

Conversations were executed sequentially within each model run and saved incrementally after each conversation to prevent data loss from interruptions. Rather than using VPNs or IP-based geolocation, location context was provided through the system prompt: ``The user is a citizen and resident of \{country\}.'' For the United States, the prompt additionally specified ``living in Texas'' to test subnational service delivery in a federal system (This subnational cue was applied only to the United States, and a robustness check adding a federal-system indicator leaves the cross-country pattern unchanged, with a negligible coefficient and the United States only 0.18 point above its EGDI-predicted chat score, so the prompt does not materially inflate the results). This explicit approach was chosen over IP simulation for two reasons: it is transparent and reproducible across models (some models may not use IP signals at all), and it captures the user experience of someone who tells an AI assistant where they are located, which is the more common interaction pattern for government service queries.

Chat mode responses were evaluated using an LLM-as-judge approach (Zheng et al. 2023). The approach allows for automated scoring across the very large assembled dataset, and avoids human bias from fatigue and repetition. On the other hand, LLM judges may share systematic blind spots with the models they evaluate and require careful quality control. To mitigate risk, we employed a cross-model judging design in which two models served as judges: DeepSeek and Gemini. Each judge evaluated all conversations except those generated by itself, preventing self-evaluation bias. The two judges differ in leniency: on the same 0-to-10 scale, Gemini as judge scores about 1.4 points higher on average than DeepSeek, consistently across both GPT-5 (8.51 versus 7.04) and Claude (7.44 versus 6.10) outputs, an offset that reflects judge behavior rather than favoritism toward any output model, consistent with documented reliability limits of LLM judges (Norman et al. 2026). Because each output model is scored by a fixed set of judges, we read within-mode score differences with this offset in mind and rely on cross-country rankings, which a constant judge offset does not affect. This design yields two independent assessments for two of the four models, providing a check on inter-judge consistency, while ensuring no model evaluates its own outputs. In total, 19,917 judge evaluations were produced. The doubly-judged subset supports a direct reliability check. Across the 6,594 responses that both judges scored, agreement on any single response is moderate: the intraclass correlation is 0.48 for absolute agreement and 0.62 for consistency, and the judges land within one point of each other on 48 percent of responses and within two points on 73 percent. Agreement is concentrated in verifiability (quadratic-weighted kappa 0.61) rather than specificity (0.37) or depth (0.27), while transparency sits at its ceiling for 98.9 percent of responses and so carries no usable agreement signal. Reliability is considerably higher at the country level, which is the unit RADAR reports: the two judges rank countries almost identically (r = 0.96, Spearman 0.96, Kendall tau 0.84) and reach a consistency intraclass correlation of 0.90. The design also balances leniency by construction. Claude and GPT-5 outputs are scored by both judges, DeepSeek outputs by Gemini and Gemini outputs by DeepSeek, so each judge carries equal weight in the four-model chat average and the leniency offset cancels exactly in the reported country score. Recomputing every country average net of the judge effect reproduces the published values to five decimal places.

Judges scored responses on the four informational legibility dimensions described in Section 2 (verifiability, specificity, depth, transparency), awarding 0--10 points total. The rubric explicitly instructs judges to penalize confident statements lacking sources (even if plausible) and to treat broken, irrelevant, or circular links as non-verifiable. Judges evaluate the quality of evidence rather than the generosity of the underlying government policy.

A deliberate design choice: the rubric assesses verifiability rather than accuracy. Establishing ground truth for 20 services across 166 countries would require extensive manual verification of government policies, fees, and procedures, resources beyond this study's scope. Verifiability (whether claims are supported by cited official sources) can be assessed from the response itself. This approach is conservative: a response citing accurate information without sources scores lower than one citing verifiable sources, even if both are correct. The tradeoff accepts some false negatives (penalizing accurate but unsourced claims) to avoid false positives (rewarding plausible-sounding fabrications).

A note on scoring scale. Each chat service score is the sum of its four dimension sub-scores (verifiability, specificity, depth, and transparency, with maxima of 3, 3, 3, and 1 respectively), placing service scores on a 0--10 scale. Per-model scores are the mean of the 20 service scores at full precision, and mode and overall averages are means of these per-model values.

All chat mode prompts were conducted in English. The questions are simple enough that it was judged that employing translators for the 100 plus languages involved, rather than using machine translation, would not have been wise. However as all translation now runs on these same models, first translating then requesting would have simply externalized an internal first step that the LLM itself will make. Some empirical results from an earlier generation of smaller models suggest that stating the initial question in a given language will skew the model to answering from specifically stored corpus materials in that language. But this would only affect results if model responses were skewing towards fabricated or incorrect English results, for which we do not see evidence (and the models we are prompting are significantly larger and more intensively post-trained for retrieval than those used in prior research). That said, the training corpus for models skewing to English is expected to exert a significant linguistic penalty on the legibility of non-Anglophone states, a hypothesis we test directly in Section 4.

\subsection{Agent operability}
Agent operability refers to whether an automated agent can reliably discover the official service entry point, navigate the portal's information architecture, and reach service-specific content rather than getting stuck on landing pages, interstitials, or broken flows. Operability is a descriptive measure of whether an agent can reach a service, not a normative assessment of whether every barrier should be removed: some barriers may reflect legitimate security or authorization controls, a distinction addressed in Section 5.3. We deliberately do not incorporate the desirability of a barrier into the operability score. Doing so would combine the measurement of agent reach with the separate policy judgment about when automated access should be permitted, obscuring the tradeoff the measure is intended to reveal. Operability can differ depending on whether agents are operating sites using their underlying structure (the ``Document Object Model'' or DOM) or their visual presentation, or, simplifying somewhat, whether the agents operate sites by reading them in code and hence as text or pointing and clicking by operating a browser and interpreting images or calling tools built for them to access.

Agent operability fails when an agent cannot reach content a human user could reach with a few clicks. The user receives no signal that the agent was blocked: only that the task did not complete. Agent operability succeeds when an agent can find what it needs to do across a wide range of tasks and any specific procedural failures are easily diagnosed and understood as substantive failures, not accidents of web page-agent interaction.

It is worth noting upfront that ``legibility'' and ``operability'' are not just different constructs, they are measured differently and have different implications for investment and policy:

A government that wants to improve informational legibility should focus on content and its distribution: publishing clear, structured, authoritative information, ensuring official sources are crawlable and indexable, using consistent terminology, keeping content current, making jurisdiction and eligibility criteria explicit. It should also focus on what is known as ``generative engine optimization'' (GEO), which involves shaping both the substance of content and its delivery to optimize its chances of being included in LLM answers (Aggarwal et al.~2024, Wan et al.~2024, Ma et al.~2025).

A government that wants to improve agent operability should focus first on front-end implementation: ensuring stable URLs and deep links, avoiding heavy reliance on client-side rendering, calibrating bot policies to allow legitimate automated access, testing cross-domain flows, providing machine-readable navigation structures, whether as schemas, tool protocols like the Model Context Protocol (MCP), or improving underlying APIs first.

Operability is measured in two different modes, corresponding to the ways agents can navigate online services. In ``DOM mode'', an agent interacts with the ``document object model'' of a website. The agent fetches the website content, including any dynamically rendered content, but does so through fetching the underlying source code and text in the page. In other words, the agent ``sees'' the site and manipulates it as if it were text. In ``Visual mode'', an agent controls a web browser and takes screenshots of sites, then interacts with the site directly through the browser.

Both DOM and Visual modes measure agent operability, but they fail in different ways. DOM mode struggles with JavaScript-heavy sites, dynamically loaded content, and non-standard markup. Visual mode struggles with ambiguous layouts, low contrast, and sites that look navigable but have broken interactive elements. The gap between them reveals something about the nature of a portal's barriers and can proxy for general usability. In both modes, we ask the agent to complete a single common service, obtaining a birth certificate, up to the point where a login or other personal credential is required. Which authority issues that certificate depends on country context. It is national in centralized systems and municipal or provincial in decentralized ones, for example municipal in the Netherlands and Bosnia and provincial in Canada. The agent is scored on the portal of whichever level is competent, so a country's operability score can reflect a municipal office rather than a national portal, a difference in administrative level that can affect the comparison, and the agent score rests on this one service rather than the twenty used for chat.

We audited the target domain each agent reached against the country's official service channel. In five countries no online service to obtain a birth certificate exists, and the agent's recorded target was therefore a non-service site: a private site in Chad and the Democratic Republic of the Congo, and a government information portal in Eritrea, North Korea, and the Republic of the Congo. In these cases the low operability scores reflect the absence of a reachable online service rather than the quality of any particular portal. Excluding them changes the operability mean by less than 0.1 point and none of the reported results, so we retain all 166 countries and mark these cases in the released data.

In the same way that we would measure a website's usability by asking humans to perform a task and then asking those same humans whether they found the task easy or difficult, using a scoring rubric to break down where and how they found it easy or difficult, we then ask the agent itself to perform the evaluation of the country-service pair with which it interacted. This is a methodological distinction between the legibility and operability mode, reflecting the different role that AI is playing in each mode, in one as provider of information, in the other as direct user.

The agents do still record detailed logs and take screenshots of the steps they take as they proceed. In both modes, we use those logs to run cross-model audits for robustness. We send the logs to another model and ask it to review the scores and suggest any adjustments. We do not blindly trust the auditor, for example ignoring minor adjustments it proposes that are not based on positive evidence, but apply adjustments where the auditing model is able to identify clear scoring errors based on the logs. The specific agent prompts along with their scoring rubrics, and the auditor model prompts are reproduced in Appendix B.

Agent modes make significantly higher demands on LLMs than do chat sessions. Instead of three small prompts with their replies, agent mode runs require elaborate contextual and system prompts to initiate, and then adding large amounts of context as either images or source code to the model interaction, along with requesting specific tool use instructions back. Even though agents are able to work autonomously for longer periods, model performance can degrade as context becomes very long.

Our primary tool to manage demands on agent context was to spawn a new sub-agent session for every country, through a script, with another agent checking on progress and alerting on general process errors or issues.

To manage the demands on the orchestrator, we batched country evaluations rather than processing all 166 countries in single runs. Batching introduces a new risk: scores might depend on timing idiosyncrasies rather than portal quality, if a country happened to fall in a ``good'' or ``bad'' agent run. To detect such drift, three reference countries were included in every batch: Estonia, the United States, and Turkmenistan. They were selected during pilot work because they occupied high, middle, and low positions in the observed score distribution, and they were used externally to check for run-to-run drift rather than as targets encoded in the scoring prompt. These reference countries appeared at varied positions within batches (start, middle, end) to test for position effects. The decision rule: if their scores varied by more than 2 points across batches, we would either re-run the batch or apply normalization adjustments, but this was not necessary once the agent prompts had been refined and modern agent harnesses used.

Initial pilot runs revealed severe ceiling saturation: agents assigned scores of 9 or 10 even when their own run logs documented substantial navigation failures. We therefore iteratively refined the evaluation prompts to strengthen evidence requirements, to require that scores reflect observed navigation failures, and to reduce positive scoring bias. The prompts were then frozen before the runs that produced the data reported in this paper, and all reported country evaluations use the same final evaluation instrument. That final prompt does not encode expected score distributions or anchor comparisons; it enforces discrimination through per-dimension evidence requirements and post-run consistency checks rather than through target score ranges.

\subsubsection{DOM mode}
DOM mode tested the same four models as chat mode: GPT-5, Claude, Gemini, and DeepSeek. Using identical models across chat and agent modes enables within-model comparison: does a model that excels at describing government services also excel at navigating government portals? The results (Section 3) reveal substantial variation, with some models performing well on one construct but poorly on the other.

DOM agents were implemented using browser-use with Playwright, enabling programmatic browser control and DOM inspection. Each country evaluation had a 90-second timeout per navigation attempt. All agent runs were executed from a single cloud region (US-Central) using a consistent egress IP pool. This external vantage point is intentional (capturing the experience of users accessing portals from outside the country) but may affect results for portals that perform differently based on geographic origin. Future work may utilize multiple cloud regions and VPNs to examine such effects. Sites that loaded slowly but eventually rendered were scored based on whatever content appeared within the timeout window. Errors were caught and countries retried up to three times before being marked as inaccessible. We distinguish between ``blocked'' (the portal actively rejected automated access, which is substantive data about AI readiness) and ``inaccessible'' (tool failure, server downtime, or network errors unrelated to bot policy). Only the latter is treated as missing data in the released dataset.

All agent runs used standard browser automation and did not attempt to defeat bot detection, CAPTCHAs, or rate limits. Portals that actively rejected automated access were recorded as blocked rather than circumvented, and each country was evaluated once with up to three retries, so the request burden on any government site was low.

\subsubsection{Visual mode}
\emph{Visual mode} instructs an automated agent to accomplish the same navigation task using only visual interpretation of the rendered page: screenshot analysis, identifying buttons and links as a human would see them. This measures operability under conditions closer to how humans (and some vision-based agents) experience websites, and captures failures that DOM-based interaction might miss (e.g., visually misleading layouts, content obscured by overlays).

For each country, agents were instructed to: discover the official government e-services portal using the country's primary official language, navigate to the main portal and examine homepage structure, locate a common service (obtaining a birth certificate), evaluate whether services can be completed online or require offline steps, and score the portal across six operability dimensions: Findability, Portal Quality, Agent Permeability, Service Access, Structured Access, and Navigation Efficiency (the full prompts are in Appendix B). The birth-certificate task was chosen because it is near-universal across countries, tends to sit at a comparable level of administrative routine, and allows evaluation up to the authentication boundary without initiating a real transaction. Five of the six dimensions capture how tractable the human-facing website is for an agent; only Structured Access captures purpose-built machine interfaces, and it is zero for most countries today (see Limitations). Agents were requested to record their justification with citations of specific screens they encountered, and specific navigation weaknesses.

Reflecting rapid developments in agent harnesses over the period of this work, our initial and final runs differed in model use and coverage substantially. Unlike DOM-based agents, which can be implemented using browser automation frameworks with any LLM backend, visual agents require tightly integrated screenshot capture, vision processing, and action generation pipelines.

At the time of initial data collection in December 2025, ChatGPT Agent Mode was the only production-ready visual agent capable of autonomous web navigation. However, by the later stages of the research, newly developed agent harnesses such as OpenClaw had substantially improved the ability to conduct such autonomous, visual-mode runs, flexibly using multiple models. Our final runs in April to May 2026 therefore used OpenClaw on a dedicated machine to conduct the evaluation runs using three of our four primary models again, with each specific evaluation also now encased in a sub-agent run for further context management and stability. Our fourth model, DeepSeek, did not allow for multimodal input via API at the time of evaluation, and so was not included. As a result, Visual mode pools three models where Chat and DOM pool four, an asymmetry that should be borne in mind for any cross-mode comparison and for the Overall composite.

\section{Results}
We evaluated 166 countries using a final cross-mode RADAR dataset combining Chat, Agent DOM, and Agent Visual scores. All scores are on a 0--10 scale. Overall RADAR scores range from 3.04 to 8.18, with a mean of 6.24 and a median of 6.27, indicating a distribution centered in the middle-to-upper part of the scale rather than one dominated by complete failure. The interquartile range runs from 5.60 to 7.00, the top 15 countries all score at least 7.52, and the bottom 15 do not rise above 4.83. Estonia ranks first overall at 8.18, followed by Norway (8.03), Italy (7.80), Cyprus (7.78), and the United Kingdom (7.77). The bottom of the distribution is occupied by North Korea (3.04), Eritrea (3.42), and Turkmenistan (4.13).

\begin{landscape}\begin{figure}[p]\centering
\includegraphics[width=\linewidth]{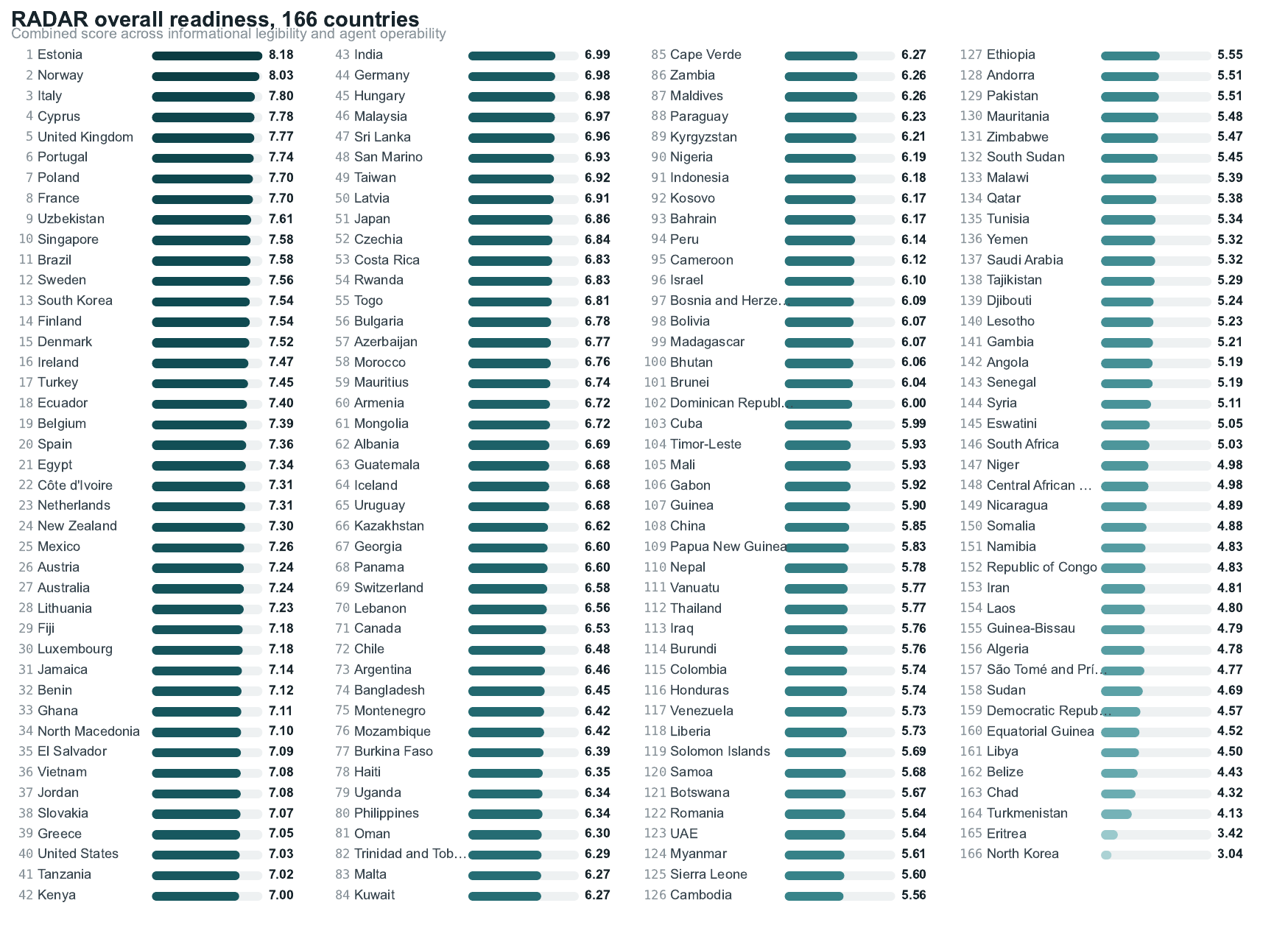}
\caption{Overall RADAR rankings and score distribution.}
\end{figure}\end{landscape}
The overall distribution is therefore better described as stratified variation than as a binary split between digitally capable and incapable states. High-scoring countries are concentrated in Northern Europe and a smaller set of other digitally mature systems, but the top of the distribution is not composed exclusively of the canonical e-government leaders: Cyprus, Uzbekistan, and Brazil all appear in the top ten overall. At the lower end, the weakest performers combine low agent operability with only middling chat legibility rather than universal collapse across all modes. This matters for interpretation: the final cross-mode surface is measuring a broad gradient of AI-mediated accessibility, not simply identifying which states have or lack online government presence.

\begin{figure}[htbp]\centering
\includegraphics[width=1.0\linewidth]{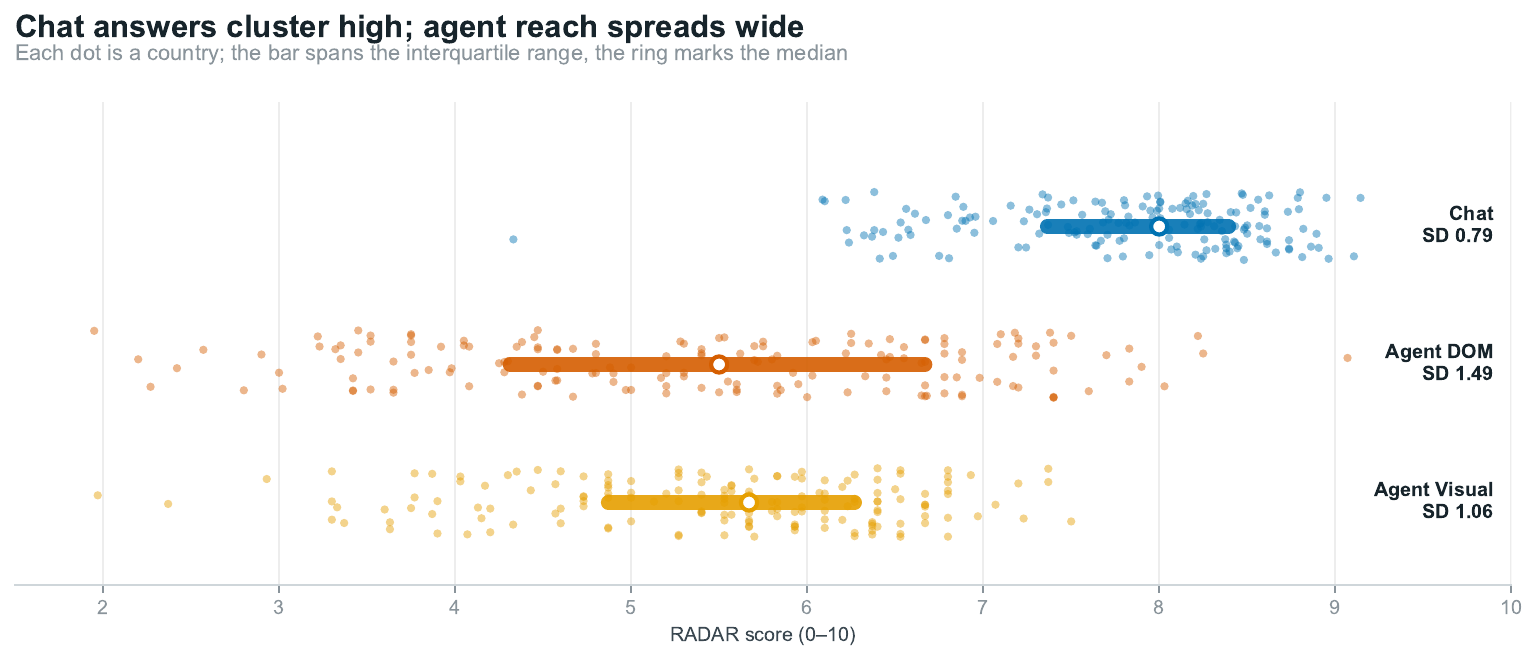}
\caption{Score distribution by mode.}
\end{figure}
The three modes differ substantially in cross-country variance. Chat is the least variable mode, with a standard deviation of 0.79 across countries, compared with 1.49 for Agent DOM and 1.06 for Agent Visual. Informational legibility is therefore much more compressed across countries than agent operability, especially in DOM mode. This suggests that large language models can provide at least moderately usable information for most countries in the sample, whereas the ability of automated agents to navigate government portals varies much more sharply.

This compression in chat scores is matched by comparatively strong within-mode consistency. Across the four chat model lines, the mean within-country spread is 1.94 points and the mean within-country standard deviation is 0.77. Pairwise correlations between chat models range from 0.43 to 0.84, with the tightest agreement between ChatGPT and Claude. The result is a mode that is not uniform, but that is substantially more stable than the agent modes both across countries and across models. In practical terms, when countries score well or poorly on chat legibility, the signal is more likely to be shared across frontier models rather than driven by one model family alone.

DOM mode behaves differently. It has the largest cross-country variance and the widest within-country cross-model spread: the mean within-country spread across the four DOM models is 2.60 points, with a mean within-country standard deviation of 1.03. Pairwise DOM-model correlations range from 0.52 to 0.78, indicating moderate but not tight agreement. This makes DOM the most country-sensitive and model-sensitive of the three surfaces. Estonia (9.07), Norway (8.25), Finland (8.22), and France (8.03) anchor the top of the DOM distribution, while Belize (1.95), Eritrea (2.20), Turkmenistan (2.27), North Korea (2.42), and Iran (2.57) sit at the bottom.

Visual mode occupies an intermediate position. Its cross-country variance is lower than DOM but higher than chat, yet its internal consistency is looser than the mode average alone might suggest. The mean within-country spread across visual model lines is 1.78 points and pairwise correlations between visual lines range only from 0.24 to 0.46. Cyprus (7.50), Uzbekistan (7.37), Belgium (7.37), Kenya (7.23), and the United Kingdom (7.20) lead the visual distribution, while Eritrea (1.97), North Korea (2.37), Chad (2.93), Algeria (3.30), and Namibia (3.30) cluster at the bottom. Visual mode therefore appears neither simply noisier nor simply easier than DOM. Instead, it captures a somewhat different and less uniformly shared view of portal accessibility.

The relationship between DOM and visual performance is best described as country-specific divergence rather than a general asymmetry. Eighty-two countries score higher on visual than DOM, eighty-three score higher on DOM than visual, and one country is effectively equal across the two modes. The mean visual-minus-DOM gap is only 0.06 points. The final consolidated dataset therefore does not support a general claim that visual navigation systematically rescues programmatically inaccessible portals. The more defensible claim is narrower. Some countries show large positive visual-over-DOM gaps, including Israel, Cambodia, Bahrain, and Peru, while others show the reverse, including Kyrgyzstan, Nigeria, New Zealand, and Estonia.

\begin{figure}[htbp]\centering
\includegraphics[width=1.0\linewidth]{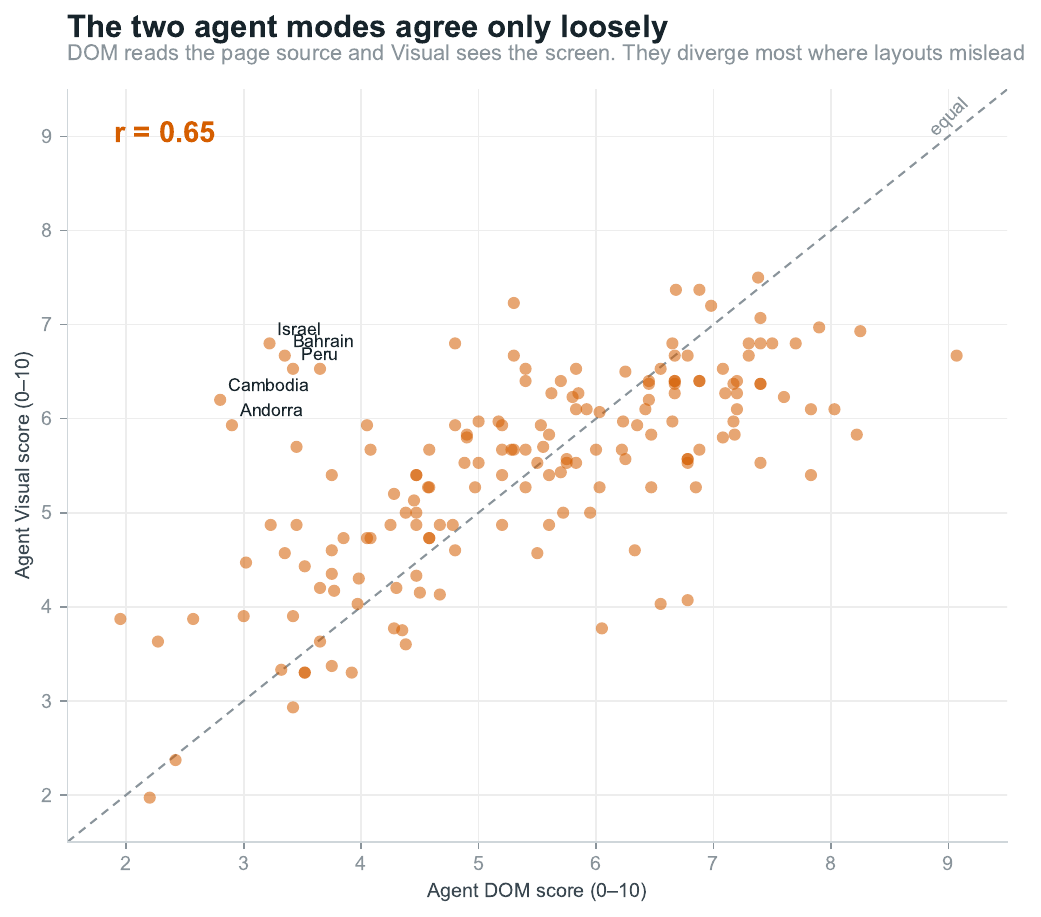}
\caption{DOM versus Visual comparison.}
\end{figure}
Chat and agent performance move together somewhat, but they are not interchangeable. The correlation between Chat and Agent DOM is 0.55, and the correlation between Chat and Agent Visual is 0.57. These are high enough to reject any view that informational legibility and agent operability are fully separate worlds, but low enough to show that the two constructs capture different constraints. A country can score strongly on chat because its services are well documented, internationally visible, or well represented in model-accessible sources, while still performing weakly in agent navigation because of portal architecture, flow design, or implementation-specific friction.

This distinction is clearest in the guidance-to-reach gap, defined here as Chat minus the Agent Operability Index, the mean of the DOM and Visual scores. Because informational legibility and agent operability are distinct constructs measured with different task and rubric structures, the magnitude of this difference should be interpreted descriptively rather than as a cardinal measure of distance between the two capabilities; the comparison is most informative in its direction and cross-country pattern. The mean gap is 2.37 points in favor of chat, and it is positive for every country in the sample: even the best-aligned cases never show agents outperforming chat. This universality is a property of the mean-based index. Under a best-channel definition that takes the higher of the two agent modes, one country, Estonia, reverses, so the finding is that chat is at least as legible as the average agent channel, not that no single channel can match it. Saudi Arabia and South Africa show the largest positive gaps, both at 5.12, followed by the UAE and Iran (each about 4.77), Belize (4.57), and Nicaragua (4.33). These are countries where frontier models can describe the service environment much more effectively than agents can operationally reach it. At the other end of the distribution, the gap never becomes negative: even the best-aligned cases, such as Côte d'Ivoire (0.49), Togo (0.83), Estonia (0.92), and Cyprus (1.01), are countries where agent performance approaches chat performance rather than surpassing it.

This gap does not widen with national wealth. The rank correlation between the guidance-to-reach gap and log GDP per capita is only $-$0.09 and is not statistically significant ($-$0.13 for the DOM component alone, also not significant), so richer countries do not systematically show larger gaps and poorer ones do not show smaller ones. The shortfall is roughly constant across the income distribution. What varies is the individual case rather than the income level: the widest gaps belong to a mix of high-income Gulf states (Saudi Arabia, the UAE) and middle-income countries, several of which run newer, more JavaScript-heavy portals (South Africa, Iran, Belize), while the best-aligned cases also span the income range. The bottleneck is therefore better read as a property of portal engineering choices, which a government at any income level can make or avoid, than as a function of national resources.

\begin{landscape}\begin{figure}[p]\centering
\includegraphics[width=\linewidth]{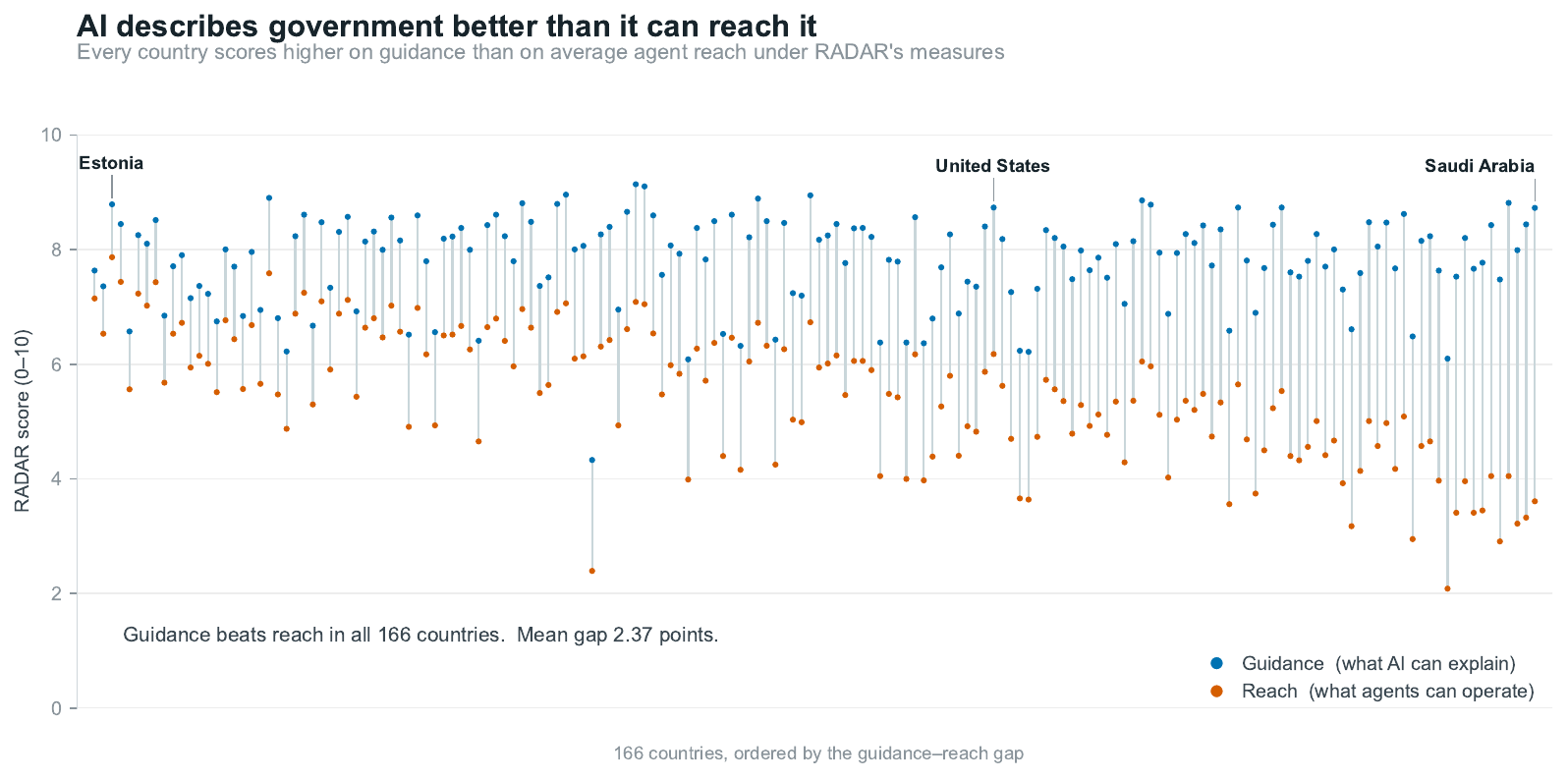}
\caption{Guidance-to-reach gap.}
\end{figure}\end{landscape}
Per-model agreement differs by mode in ways that bear on how much weight each mode can carry. Chat models correlate most tightly (pairwise r up to 0.84, ChatGPT and Claude closest), DOM models agree moderately, and Visual models agree least (pairwise r as low as 0.24). DeepSeek scores far higher on chat than on agent navigation and is absent from Visual mode entirely, but part of that contrast is an artifact of the judging design rather than a property of the model. DeepSeek's chat responses are scored only by the more lenient judge and Gemini's only by the stricter one, so per-model chat levels are not directly comparable. Once the 1.4-point judge offset is netted out, DeepSeek's chat advantage over GPT-5 and Claude falls from 1.13 to 0.42 points, and its chat-to-agent gap (3.39 points) sits close to Gemini's (3.02) rather than standing apart from the field. Country scores are unaffected, because each judge carries equal weight in the four-model average. Beyond its role as a reliability check, cross-model agreement may itself carry information: where independent models converge on the same guidance, that convergence is a plausible (if imperfect) proxy for accuracy, since fabricated specifics are less likely to be shared across systems than correct ones. We treat consistency here as a measurement-quality signal, but it is a candidate metric in its own right.

\begin{landscape}\begin{figure}[p]\centering
\includegraphics[width=\linewidth]{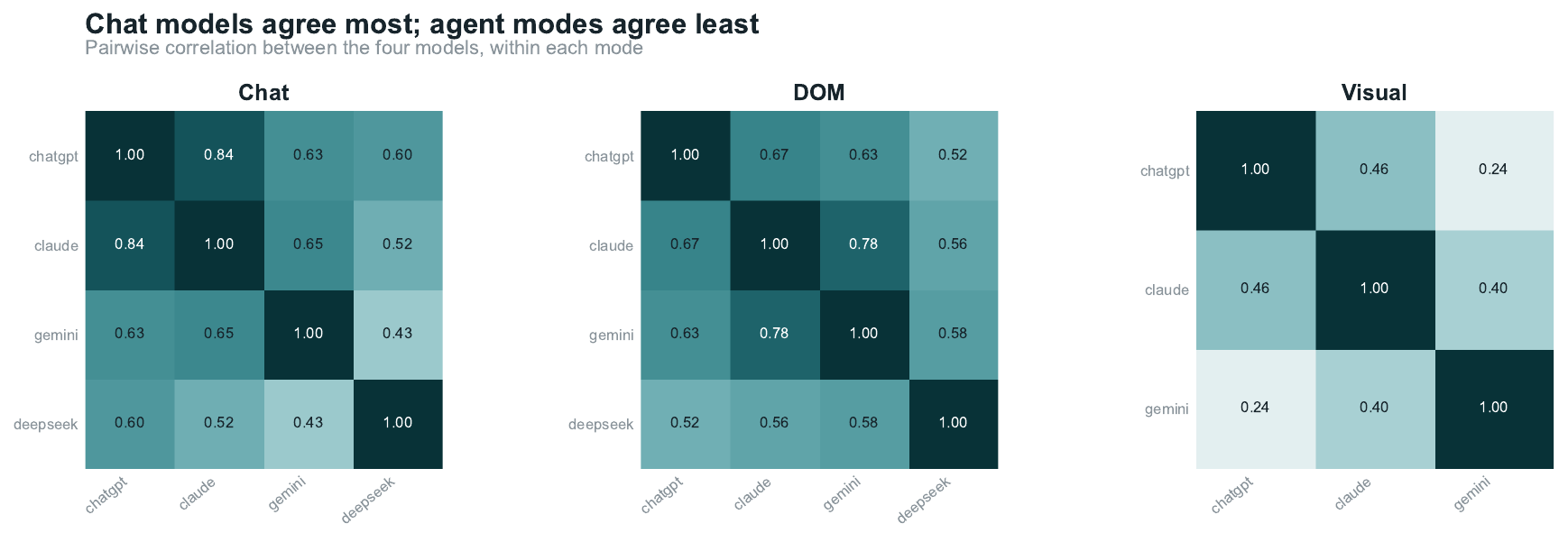}
\caption{Per-model agreement and cross-model spread.}
\end{figure}\end{landscape}
A final substantive result is that conventional digital-government covariates predict chat performance much better than they predict agent performance. Matching the RADAR sample to the UN E-Government Survey 2024 yields 164 countries with EGDI and OSI coverage, all except Taiwan and Kosovo, which the survey does not cover. EGDI alone explains 69.1 percent of cross-country variance in Chat scores, but only 23.9 percent of DOM variance and 27.4 percent of Visual variance. OSI produces the same pattern even more strongly: it explains 76.8 percent of Chat variance, but only 30.3 percent of DOM variance and 32.0 percent of Visual variance. The information large language models can provide about a country therefore tracks conventional digital-government development relatively closely, but agent performance depends much more heavily on additional portal-level implementation choices not captured well by standard e-government indices.

The cross-country variation in chat legibility is also concentrated in particular services rather than spread evenly. Breaking the chat scores down by service and comparing high-income to low-income countries, healthcare is the most income-stratified domain: the gap for setting up a GP appointment is 2.90 points and for diagnostic testing 2.78 points, roughly double the gap for filing income taxes (1.18 points). Tax procedures are well described across the income distribution, plausibly because they are relatively standardized and heavily documented, whereas primary-care booking is fragmented across providers and far less uniformly digitized, especially where no single national portal exists. The hardest services in absolute terms are the locally administered, often local-language ones: GP appointments (mean 6.90), marriage certificates (7.08), and ID renewal (7.38) score lowest, while training, government job listings, and general portal discovery score highest.

\subsection{Peer-relative performance: who beats their income and digital development}
To separate genuine over- and under-performance from the income and development gradient, we matched RADAR to two external references on ISO3 country codes: World Bank region and income classification, and the UN E-Government Survey 2024 (EGDI, OSI, and the underlying sub-indices). We then computed each country's residual after regressing its score on EGDI. Positive residuals mark countries that perform better than their digital development predicts. Negative residuals mark the reverse.

The over-performers on overall RADAR are concentrated among lower-income, lower-EGDI states, several in West and East Africa: Côte d'Ivoire, Togo, Benin, Tanzania, and Ghana all clear their EGDI-predicted scores by a wide margin (Ghana by +0.86). The under-performers are high-EGDI states whose AI accessibility lags their digital infrastructure, including several Gulf states and information-restricted regimes (Saudi Arabia, the UAE, Iran, Turkmenistan), alongside South Africa. This is the external-benchmark version of the ``sophistication penalty'' discussed below: digital infrastructure does not translate automatically into AI accessibility, and some governments with modest infrastructure are markedly more accessible than their resources would predict.

\begin{figure}[htbp]\centering
\includegraphics[width=1.0\linewidth]{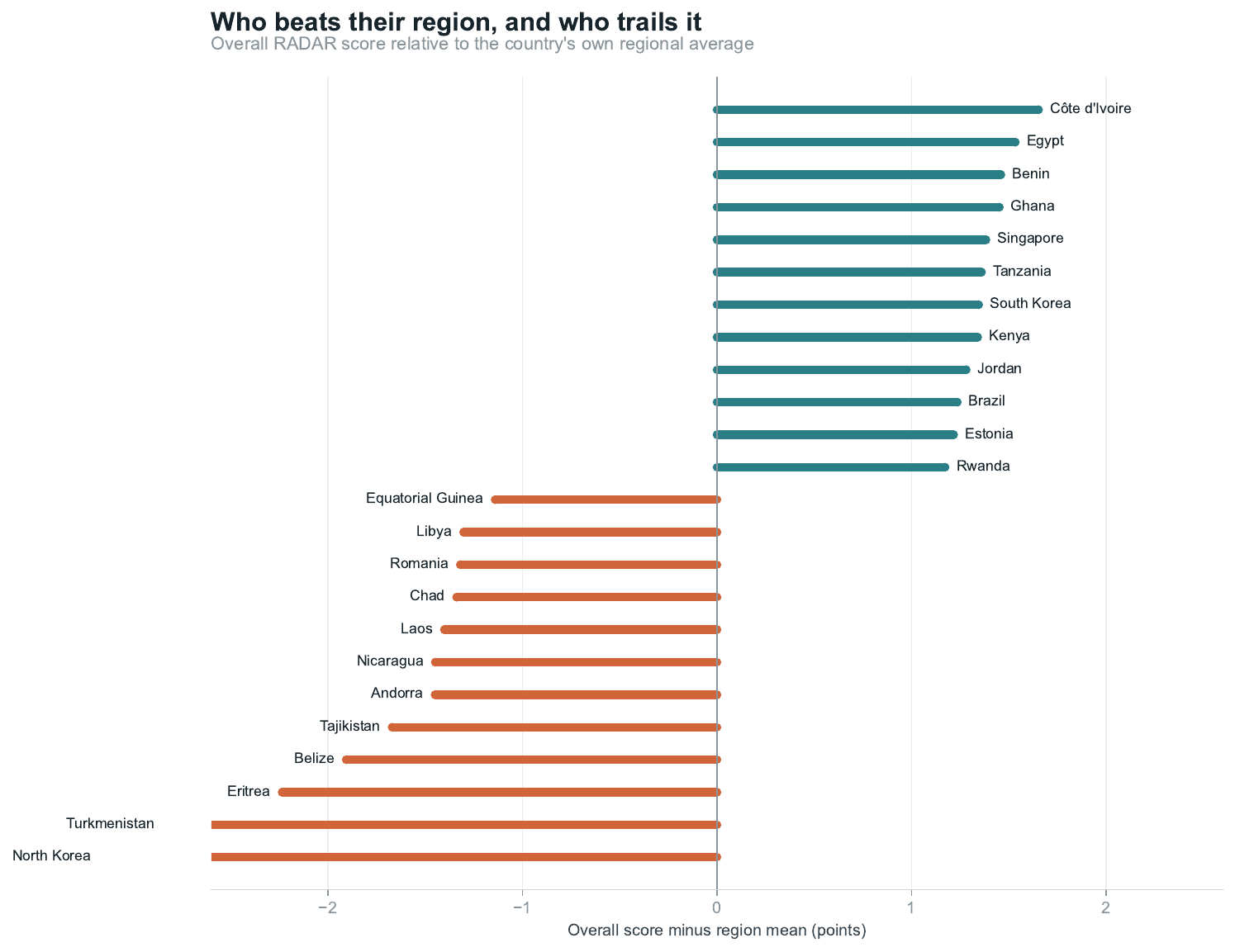}
\caption{Within-region over- and under-performers (Overall).}
\end{figure}
\begin{landscape}\begin{figure}[p]\centering
\includegraphics[width=\linewidth]{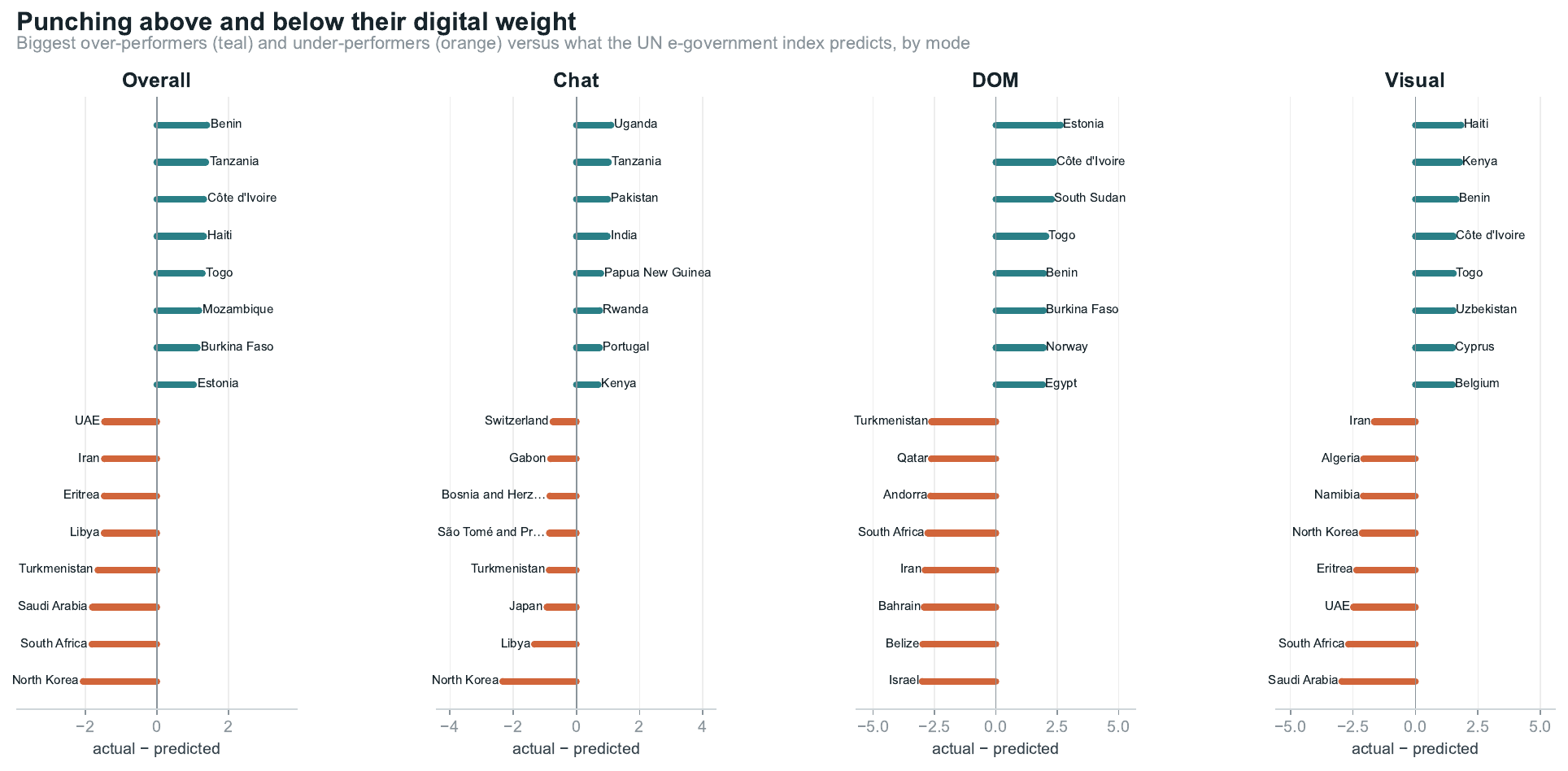}
\caption{EGDI residuals: over- and under-performers relative to digital development.}
\end{figure}\end{landscape}
\section{What predicts readiness: language, web presence, and the two channels}
The result that conventional indices predict chat far better than agent navigation raises a sharper question: what, specifically, is associated with each construct? RADAR's two failure modes map onto a distinction the AI literature draws between parametric knowledge, what a model learned during training and recalls at inference, and retrieved knowledge, what it fetches in real time through search and tool use. Chat legibility, especially without retrieval, leans on parametric knowledge. Agent operability is inherently a real-time, retrieval-and-navigation task. If the two constructs draw on different information channels, their observed correlates should separate along this line. They do.

Language coverage and the parametric channel. We proxied each country's representation in web-scale corpora using Common Crawl, a large web corpus widely used in LLM pre-training. For each country we computed the share of all Common Crawl pages written in its primary official language (crawl CC-MAIN-2026-25), distinguishing the dominant administrative or lingua-franca language (English, French, Arabic, and so on) from the indigenous national language where these differ. Net of EGDI, administrative-language coverage is associated with Chat legibility (partial r = 0.33) and explains 11.2 percent of the cross-country variation left unexplained by EGDI, but contributes essentially nothing to agent operability (partial r of 0.03 and 0.00 for DOM and Visual). The indigenous-language variant predicts nothing in any mode. In other words, chat legibility rides on whether a country's content exists in a globally well-represented, high-resource language, not on whether it is published in the local tongue: a government publishing in Hausa or Quechua gains no parametric benefit, while the same content in English or French does.

Web presence and the retrieval channel. Both measures come from Common Crawl but capture different things. The first captures language representation in web-scale corpora: the share of Common Crawl pages written in a country's main administrative language. The second shifts from language to country-specific web presence. We count the pages captured by Common Crawl under each country's national internet domain, such as .br for Brazil or .rs for Serbia, and use that volume as a proxy for the amount of country-specific web content visible to web crawlers. Net of EGDI, national web presence is associated with agent operability (partial r = 0.30 for the Agent Operability Index, the mean of DOM and Visual; 0.27 and 0.24 for DOM and Visual separately) and explains 9.1 percent of the variation left unexplained by EGDI on the 59-country ccTLD sample, but is not significantly associated with Chat. This measure is available for only 59 of the 166 countries, because Common Crawl publishes top-level-domain counts only above a 0.05 percent floor, so the ccTLD result is restricted to countries with substantial national web presence and is selected on the predictor itself; it should not be read as a global estimate for the full sample. On that subsample the DOM association is significant (t of about 2.1) while the Visual association is not (t of about 1.9, p of about 0.07), reaching conventional significance only with the population control below. A crawlable national web presence is what agents draw on in real time; it does nothing for a model recalling service details it never indexed. This is not a country-size artifact: adding log population as a control does not weaken the relationship but strengthens it (partial r rises to 0.39 for DOM and 0.30 for Visual), a suppression pattern in which population correlates positively with raw page counts yet negatively with operability at fixed digital development, so that holding size constant reveals a cleaner web-presence effect.

Together these show a divergence in observed correlates: language coverage is associated with chat and not operability, while web presence is associated with operability and not chat. This pattern is consistent with the two modes depending more heavily on different information channels, parametric knowledge for chat and retrieved knowledge for agent operability, though the analysis is observational and does not isolate either channel experimentally.

\textbf{The language effect is specific, and corroborated independently.} A complementary analysis regressed each mode's score on an English-language-portal indicator while controlling for log GDP per capita, internet penetration, government effectiveness, and log population, with robust standard errors. English-portal status independently predicts Chat (coefficient +0.34, p < 0.001) but neither DOM (+0.11, n.s.) nor Visual (-0.03, n.s.). Two independent operationalizations, a continuous corpus-coverage measure and a controlled binary indicator, point the same way: language operates through the informational, parametric channel, not the navigational one.

This parametric gap leaves a concrete signature inside the chat scores. The chat rubric scores four things: how specific an answer is, how much depth it provides, how transparent it is about uncertainty, and whether it cites the correct, working URL (Verifiability). Three of the four are near-saturated, with even low-income countries scoring well on specificity, depth, and transparency. The only sub-dimension that meaningfully separates countries is Verifiability, which falls from 2.04 in high-income countries to 1.04 in low-income ones, a full rubric step, while the others differ by 0.40 points or less. Models therefore produce specific, detailed, well-hedged answers for almost every country, but get the official link wrong precisely where that country's government web presence is thin and under-indexed in training data. The informational failure is not vagueness but misdirection to a wrong or dead address, which is exactly the failure a retrieval step is positioned to correct.

\begin{figure}[htbp]\centering
\includegraphics[width=1.0\linewidth]{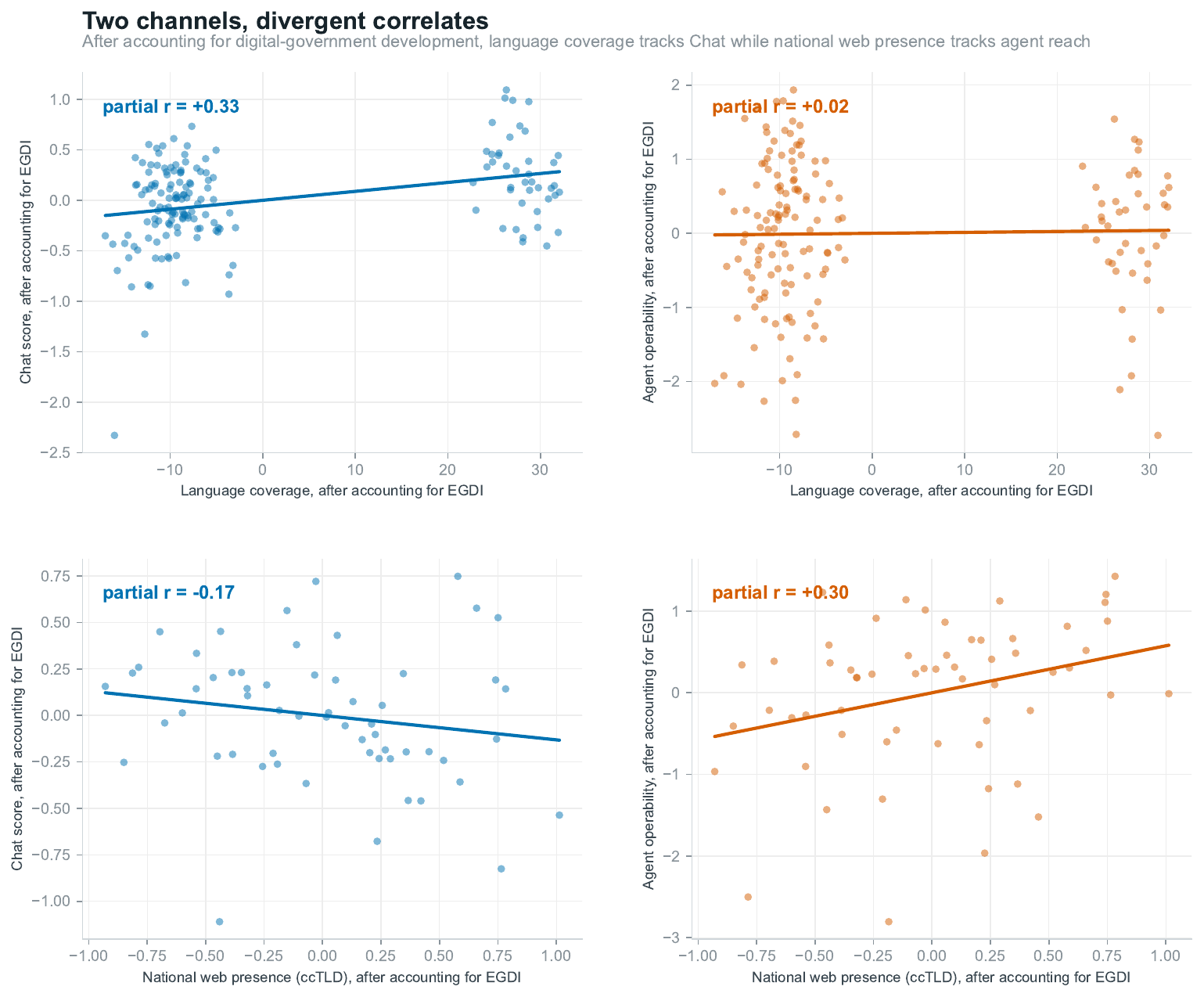}
\caption{Common Crawl coverage versus RADAR mode, net of digital-government development: language coverage tracks Chat, national web presence tracks operability. Each panel shows the relationship after removing the variation associated with EGDI. The Operability panels use the Agent Operability Index, the mean of the DOM and Visual scores; its partial correlations therefore differ slightly from the separate DOM and Visual values reported in the text, because averaging the two agent modes reduces mode-specific noise.}
\end{figure}
Three cautions bound these results. First, the effects are modest in magnitude: net of EGDI, each adds only a few percentage points of explained variance. Conventional digital development still explains most of the cross-country variation. Second, a dedicated government-domain coverage measure (built and validated for all 166 countries) added essentially nothing beyond EGDI, indicating that the web-presence signal reflects general national web maturity rather than government web specifically. Third, because the chat protocol was conducted in English, the language effect may partly reflect content-query language matching (a country's content being in the same language as the query) as much as training-corpus composition. The continuous coverage measure mitigates but does not fully resolve this. With those bounds noted, both halves of this divergence in correlates are robust to development and size: language coverage predicts chat legibility and not agent operability, while national web presence predicts agent operability, on the 59-country subsample where it can be measured, and not chat.

A parallel contrast appears among institutional predictors. Regressing each mode on standardized country characteristics (log GDP per capita, internet penetration, the World Bank's Government Effectiveness indicator from the 2024 Worldwide Governance Indicators, and EGDI), digital-investment indices dominate chat: EGDI alone explains about 69 percent of chat variance, ahead of government effectiveness (roughly 52 percent), internet penetration (48 percent), and income (44 percent). For agent operability the spread compresses and the order changes: no single predictor explains more than about a quarter of DOM variance, and government effectiveness edges past EGDI (26 versus 24 percent), with income and connectivity further behind.

This fits the two-channel reading. Informational legibility tracks how much a country has invested in digital-government inputs, which is what EGDI is built to measure, whereas operability depends more on how that investment is implemented and administered, which a governance-quality measure proxies better than an inputs index. We read this as suggestive rather than decisive: the operability margins are small, the visual mode does not reproduce the reordering (EGDI and government effectiveness are level there), and in a combined model EGDI remains the largest single coefficient even for DOM. The defensible claim is comparative, that building portals agents can actually use looks more like a public-administration problem than an EGDI-checklist one, not that any single index governs operability the way EGDI governs chat.

The policy reading of this section is taken up in Section 5: the two channels are fixable by different actors on different timescales.

\section{Policy Implications}
AI mediation is turning government's online presence into a de facto service channel. Citizens who ask chatbots about business registration or passport renewal are not navigating government portals: they are relying on systems that retrieve, synthesize, and present information on their behalf. LLMs not only provide access to these services, but advice on how and whether to use them. Whether that information is accurate, current, and grounded in authoritative sources depends substantially on design choices governments control.

The RADAR results suggest that these choices are not being made deliberately. Some traditional digital government leaders score poorly on AI accessibility, while other countries with minimal digital infrastructure may outperform them. This is not because simpler portals offer better services. It is because architectural complexity creates barriers that are invisible to human users but determinative for automated systems. The policy challenge is to make these tradeoffs visible and manageable. We frame the government's response as sovereign legibility, the deliberate work of making authoritative content discoverable, structured, and interpretable by AI systems, and the sections below are its instruments.

\subsection{Recognize AI-mediated access as a service channel}
The first step is recognition: AI-mediated access should join the web-traffic, uptime, mobile-responsiveness, and accessibility checks that governments already run, not as a speculative future concern but as a current reality.

A practical first step is to institutionalize ``AI readiness checks'' alongside existing quality checks, using lightweight test suites that can be repeated after major content or platform changes. The RADAR methodology, or simplified versions of it, could support this. Unlike lengthy and costly survey-based indices that rely on government self-reporting or expert assessment of portal features, the protocol is automated and replicable: the 166-country assessment reported here was completed in weeks at a fraction of the cost of traditional e-government measurement. This makes ongoing monitoring feasible rather than a one-off exercise. This is also where administrative capacity matters: government effectiveness predicts agent operability roughly as strongly as digital-government investment (Section 4), consistent with day-to-day implementation capacity, not only infrastructure, shaping whether services remain reachable.

Sites should also put in place the infrastructure to distinguish bad actors (``bots'') from authorized, permissioned agent use. This is where MCP tools, particularly including the use of OAuth protocols in such tool use, can be useful. Of course, bad actors will attempt to use MCP just as they attempt to use published APIs, and there is sometimes a tendency to ``MCP-wash'', or pretend that simple wrapping of an API in an MCP tool layer is on its own enough to make a site agent ready. One avenue to explore may be to have registries of certified agents, and within those the ability for residents to authorize agents to act on their behalf. It will be important to make any such effort a promoter of uses rather than a layer of veto licensing, with clear and first principles based and automated reviews, rather than discretionary ones, as well as potential subsidies to allow for equal access. The private sector is also still learning how to navigate the tension between keeping out malicious bots and allowing permissioned agents. Governments should therefore continue learning across sectors and countries as approaches emerge that balance security with innovation and the productive use of this technology.

\subsection{Improve informational legibility through structured, authoritative content}
To reduce wrong or untraceable guidance, governments should make it easier for AI systems to ground answers in authoritative sources. The priority is procedural truth: eligibility rules, document requirements, fees, timelines, and current official entry points.

Three interventions tend to matter across contexts:

\emph{Publish canonical service pages:} Each service should have a stable, clearly scoped page that is the authoritative source for that jurisdiction. Explicit ``last updated'' metadata helps both human users and AI systems assess currency.

\emph{Structure the guidance:} Critical requirements should not be buried in PDFs, scattered across announcements, or accessible only through multi-step navigation. The goal is not sophisticated presentation but clear, extractable content.

\emph{Add machine-readable metadata:} Schema.org markup, structured data, and consistent taxonomy help automated systems interpret content reliably. The UK's use of schema.org to make COVID-19 guidance interpretable by search engines and voice assistants illustrates the approach (Marshall and Hughesdon 2020). Emerging conventions aimed specifically at AI consumption, such as an llms.txt file that points models to a site's authoritative, machine-readable content, are low-cost to adopt and worth piloting as they gain traction.

\emph{Consciously work on LLM-relevant distribution.} GEO promises in the private sector to have LLMs cite brands more often. The techniques of GEO are much debated. Because models draw heavily on a small set of high-pickup public sources, governments should track which third-party channels frontier models cite and ensure their authoritative content is present and correctly attributed there (Kumar 2026).

These are largely no-regret investments. The same changes that improve AI legibility typically improve SEO, accessibility, and human usability. For governments building new systems, they add minimal cost. For governments with complex existing portals, the investment is greater but compounds over time. Japan's Digital Agency has moved furthest in this direction, explicitly reframing government disclosure as preparation for AI training data, with structured formats designed for machine ingestion rather than human reading alone (Digital Agency of Japan 2025). The initiative is recent, and its effects are more likely to appear in future crawls and training runs rather than in current scores: Japan ranks 51st on RADAR here, a reminder that structured-content strategies pay off on the timescale of model training cycles, not immediately.

Two practical points follow from the finding in Section 4 that chat legibility and agent operability run on different channels. First, legibility depends substantially on what models learned during training, which is largely fixed and skewed toward high-resource languages. A government cannot change that quickly, but it can shape what models retrieve at query time, and the shift toward retrieval-augmented systems is the lever: clear, current, authoritative content can be surfaced now, without waiting for the next training run. Second, creating good content is not enough without distributing it. Models lean heavily on a small set of high-pickup sources (Reddit prominent among them), and training and retrieval pipelines index new government pages faster and more completely when the same content also surfaces in the places those pipelines favor. What moves chat legibility is a deliberate publishing-and-distribution strategy, not only a better portal.

\subsection{Improve agent operability by reducing invisible navigation friction without compromising security}
A low operability score identifies friction encountered by agents. It does not, by itself, establish that the friction is undesirable or should be removed. If automated agents become routine operational intermediaries, whether for citizens using AI assistants or businesses using RPA for regulatory compliance, governments will need to manage a new tradeoff: protecting sites against harmful automation while not blocking legitimate delegated access. This tradeoff is sharper for government than for the commercial web from which these design norms emerge, since public portals are higher-value targets and several of the changes below reduce friction for legitimate and malicious automation alike. The recommendations therefore apply to the information layer RADAR actually probes (eligibility, requirements, fees, locations) and not to transactional surfaces (authentication, payment, identity verification), where access friction is often doing exactly the job it should and a low operability score may reflect appropriate protection rather than a defect. Here, we highlight architectural patterns associated with agent failure:

\emph{Client-side rendering:} Portals that depend heavily on JavaScript to render navigation and content are systematically harder for DOM-based agents to traverse. Agents can still parse such pages, but at significantly higher cost and likelihood of time-outs. Where essential content and navigation can be delivered in initial HTML, even if subsequent content is rendered dynamically, agent operability improves.

\emph{Unstable URLs and weak deep-linking:} Agents that cannot link directly to service-specific pages get stuck on landing pages or generic portals. Stable, predictable URL structures support both programmatic access and human bookmarking.

\emph{Aggressive bot controls:} Some blocking is appropriate, as governments face real threats from scraping, credential stuffing, and denial-of-service attacks. But blanket bot-blocking also blocks beneficial automation. The policy goal is calibrated controls: rate limiting rather than blocking, risk-based challenges triggered by behavior rather than automation detection, and potentially allowlisting known agent frameworks or user-agents. Means to license agents, including to allow citizens to grant such agents digital powers of attorney, will likely be the best way to strike such a balance in the medium to long term.

The DOM-versus-Visual diagnostic helps governments understand which mode of access is failing. A large positive gap (Visual outperforming DOM) suggests that human-like navigation works but programmatic access is blocked, often fixable through architecture changes. A negative gap (DOM outperforming Visual) suggests confusing visual presentation despite clean underlying structure, a different problem requiring different interventions. Across the full sample the two modes are on average balanced (mean difference 0.06 points), so the diagnostic is best applied country by country rather than as a global claim that one mode systematically beats the other. Where the blocked surface is transactional, a positive gap may likewise record bot controls working as intended rather than a barrier to remove.

\subsection{Offer agent-friendly interfaces for high-value information flows}
For some use cases, the cleanest path is not to make websites more navigable but to make government information directly consumable through standardized interfaces.

Two patterns are emerging:

\emph{APIs as a primary surface for agent access:} Government information and service systems that expose consistent, documented APIs provide a stable interface for government and third-party agents. For public information, these interfaces can make authoritative guidance on eligibility, requirements, exceptions, fees, locations, hours, and procedures directly accessible without requiring agents to scrape human-oriented websites. For personalized or transactional services, the same architectural approach can support controlled access subject to authentication, consent, and authorization. In both cases, documented interfaces make interactions more reliable, auditable, and governable.

\emph{Protocol layers that make government interfaces usable by agents:} Tool protocols and wrappers can allow AI systems to invoke government APIs and other authoritative endpoints through standardized interfaces. The U.S. Government Publishing Office has launched a public preview of an officially supported GovInfo MCP server, giving LLMs and agents access to current content from its certified digital repository alongside the existing website and API. India has similarly introduced an MCP server connecting AI tools to verified official statistics through the national e-Sankhyiki portal, expanding from seven statistical products in its beta version to 21. These remain information-retrieval systems rather than transactional public services, but they show governments beginning to design authoritative machine-facing interfaces directly (U.S. Government Publishing Office 2026; TaxTMI 2026).

These interfaces are not equally appropriate for every layer of a service. Public information about eligibility, requirements, exceptions, fees, locations, hours, and procedures can often be exposed openly. Personalized determinations, applications, payments, and legally consequential actions generally require authentication, consent, and authorization.

The boundary between public information and authenticated services is itself an architectural choice. Governments must decide how much service intelligence to make available before a user's identity is known. Rich public information on eligibility rules, requirements, exceptions, fees, and procedures can allow an agent to provide useful contextual guidance before authentication, reserving identity checks for personalized determinations and authorized actions. Placing too little information on the public side of this boundary leaves agents unable to help until credentials are supplied; exposing personal or sensitive information creates the opposite risk. Agent-mediated services therefore raise a new design question: how much useful guidance should be available before identity is established, and what should become accessible only after authentication?

Beyond exposing authoritative information, governments and third parties can operate agents grounded in it. Governments can index their own content and expose it through retrieval-augmented assistants, documented APIs, or official MCP endpoints. Third parties can build user-facing services on top of those authoritative sources, subject to the relevant access, security, and authorization rules. By maintaining authoritative sources and verifiable provenance signals, the state can improve the substance and traceability of answers without needing to operate every interface through which citizens interact with public services. This also mitigates two risks associated with generic agents: misinformation, when an agent improvises or relies on unofficial sources, and impersonation, when users cannot verify whether an answer or action carries official authority. Official domains, signed responses, and verifiable government endpoints can help preserve that distinction. Access to an authoritative source does not itself authorize an agent to transact or speak on the state's behalf. Those permissions require separate identity, consent, and authorization controls. The falling cost of building these systems makes them increasingly feasible for governments, civic organizations, and private intermediaries alike.

\subsection{Differentiate strategy by starting position}
Starting position dictates strategy, and RADAR sorts governments into four situations. High-capacity states with sophisticated portals face a retrofit problem: systems built for human users, with heavy client-side rendering, complex session management, and aggressive bot control, are the hardest for agents to reach, so the fix is architectural rather than aspirational and often means a parallel agent-friendly access layer rather than a rebuild. A smaller group of high-capacity states has already avoided that trap. Estonia, France, and the United States score well on AI accessibility because early bets on standards, interoperable systems, and API-first design compound, and their choices show what agent-accessible design can look like. Lower-capacity governments building new systems have the opposite advantage: simpler architecture can be more machine-legible, and Ghana, which clears its EGDI-predicted score by a wide margin (Section 3.1), shows what that can look like, even if it is not yet proof of a deliberate strategy. Governments in restrictive information environments sit at the floor of RADAR by choice, not incapacity, and for them a low score records a decision about information openness that no architectural change will move.

\subsection{Use RADAR diagnostically, not only as a ranking exercise}
Country comparisons are useful, but the main value of the RADAR framework lies in separating failure modes rather than the composite league table alone. A low informational legibility score points to content and metadata problems, generally fixable through publishing practices, structured data, and content maintenance. A low agent operability score identifies navigation, rendering, or access barriers encountered by agents. The first step is to diagnose their source: some reflect avoidable implementation friction, while others may result from deliberate security or authorization controls. Large gaps between Chat and Agent Operability scores identify where that diagnosis is most urgent, particularly because users may receive no visible indication that agent access has failed. The appropriate response may be architectural redesign, calibrated automated access, or retention of a legitimate control. The composite RADAR score is useful for country-level comparisons, but the diagnostic value lies in the components and the gaps between them.

\subsection{Limitations}
Several limitations bound these results. The chat rubric scores verifiability rather than ground-truth accuracy, so a correct but unsourced answer is penalized. All chat prompts were run in English, which leaves the language finding partly confounded with query-language matching. Chat legibility is discriminated almost entirely by the Verifiability sub-dimension, and two of the six operability dimensions carry little cross-country signal, Structured Access is zero for 84 percent of observations and Findability is at its ceiling for 93 percent, so both constructs rest on fewer effective dimensions than the rubric implies. The two LLM judges differ in leniency by about 1.4 points, and Gemini and DeepSeek chat outputs are each scored by a single judge, so those scores carry no inter-judge check. On the 6,594 responses that both judges scored, agreement on a single response is only moderate (intraclass correlation 0.48 for absolute agreement and 0.62 for consistency), so an individual chat score is a noisy measurement. The country averages RADAR reports are considerably more robust (consistency intraclass correlation 0.90, judge-to-judge correlation r = 0.96), and because each judge carries equal weight in the four-model chat average, the leniency offset cancels in the reported country score. Per-model chat levels, by contrast, should not be compared directly.

Agent operability is measured on a single service, obtaining a birth certificate, so a country's score reflects the accessibility of that service's portal and may not generalize to its broader portal estate. On the agent side, all runs used a single cloud region and a fixed egress IP, so portals that serve different content by geographic origin are seen only from outside the country, and Visual mode pools three models where Chat and DOM pool four. The web-presence measure is available for only 59 of the 166 countries and is selected on the predictor, so the operability half of the language-versus-web-presence contrast is the less secure of the two. The reported correlations are cross-country and observational, carry no multiplicity adjustment, and should be read as associations rather than mechanisms.

\section{Conclusion}
For a decade governments have invested in experience for human visitors, in responsive design, intuitive navigation, and accessible interfaces. RADAR measures a second audience that arrived without notice: the AI systems that now read government content and act on it for citizens. The two audiences do not want the same things, and the gap between them is not small. Across all 166 countries, frontier models describe public services better than agents can reach them, by an average of 2.37 points on a 0 to 10 scale, and under the mean-based operability index it does not reverse. Saudi Arabia and South Africa show the largest measured gaps (Saudi Arabia: Chat 8.73, operability 3.61; South Africa: Chat 8.44, operability 3.33). As the methodology notes (Section 2.2.1), however, agent runs used a single geographic vantage point, so country-level gaps may partly reflect portals behaving differently by geographic origin rather than portal design alone.

Legibility and operability are distinct, not two labels for one thing. Chat correlates only moderately with agent navigation (r = 0.55 with DOM, 0.57 with Visual), so a government can be well described and hard to reach at once. The two failures need different owners. Content teams fix legibility through canonical pages, structured metadata, and stable official sources. Platform teams fix operability through stable URLs, server-rendered fallbacks, and bot policies calibrated to admit permissioned agents. Section 4 shows the two modes diverge in their observed correlates as clearly as the fixes differ: chat legibility is associated with how well a country's language is represented in training corpora, a slow variable largely fixed once a model is trained, while operability is associated with crawlable national web presence, a variable governments can change now.

Traditional digital-government measures predict informational legibility far better than agent operability, and where operability falls short, the reason is specific rather than sweeping. The countries that underperform their infrastructure tend to run newer, more JavaScript-heavy portals with aggressive bot control, so the sophistication penalty, where it operates, is a property of particular engineering choices, not a smooth function of wealth. No portal-architecture variable is measured here, so this remains a hypothesis about mechanism rather than a tested relationship. The mirror image is the potential leapfrog case. Ghana over-performs relative to its digital development (Section 3.1), a pattern consistent with the possibility that simpler architecture can sometimes be easier for machines to navigate. A government building now can test that possibility deliberately rather than reproduce complexity by default.

This points to what we call sovereign legibility, the deliberate effort by a government to make its authoritative content discoverable, structured, and interpretable by AI systems rather than leaving external actors to represent it. It is a no-regret strategy, not a guarantee. Models still hallucinate and still prefer unofficial sources, but the same machine-readable content at stable URLs that raises AI accessibility also improves search ranking and human usability, and it is available to resource-constrained governments that cannot change what a model absorbed in training. The choice each government now faces is whether to shape how AI represents its services or to accept whatever representation emerges by default.

RADAR measures moving targets, and several extensions would sharpen it as capabilities and portals change. Re-running the chat protocol in each country's official language would test directly whether the Anglophone advantage is a corpus effect or an artifact of English queries, and manual ground-truth verification on a stratified sample would document the error types the current verifiability rubric cannot see. These and further directions, including prompt-robustness checks, rubric refinement, and longitudinal tracking, are presented in Appendix A. Because the assessment is cheap enough to repeat, AI accessibility becomes a standing metric rather than a one-time audit. Governments do not get to decide whether AI systems read and act on their services, only whether the representation those systems produce is one they shaped or one they inherited.

\section{Data and code availability}
The RADAR dataset, including country-level scores, per-service and per-model judge scores, and evaluation transcripts, will be released under a Creative Commons Attribution 4.0 license, with the analysis and figure code under an MIT license, at a public repository with a citable DOI. Transcripts will be screened to remove any egress IP addresses, internal paths, and incidental personal data before release. The external reference data used for the covariate analysis (the United Nations E-Government Survey 2024, World Bank income and region classifications, the Worldwide Governance Indicators, and Common Crawl coverage measures) are available from their original sources.

\section{References}
Aggarwal, Pranjal, Vishvak Murahari, Tanmay Rajpurohit, Ashwin Kalyan, Karthik Narasimhan, and Ameet Deshpande. 2024. ``GEO: Generative Engine Optimization.'' In \emph{Proceedings of the 30th ACM SIGKDD Conference on Knowledge Discovery and Data Mining}, 5-16.

Chapekis, Athena, and Anna Lieb. 2025. ``Google Users Are Less Likely to Click on Links When an AI Summary Appears in the Results.'' Pew Research Center, July 22, 2025.

Chatterji, Aaron, Thomas Cunningham, David J. Deming, Zoe Hitzig, Christopher Ong, Carl Yan Shan, and Kevin Wadman. 2025. \emph{How People Use ChatGPT}. NBER Working Paper No.~w34255. National Bureau of Economic Research.

Deng, Xiang, Yu Gu, Boyuan Zheng, Shijie Chen, Sam Stevens, Boshi Wang, Huan Sun, and Yu Su. 2023. ``Mind2Web: Towards a Generalist Agent for the Web.'' \emph{Advances in Neural Information Processing Systems} 36: 28091-28114.

Digital Agency of Japan. 2025. \emph{Research on Converting Government-Held Data into AI Training Data: Final Report}. Tokyo: Digital Agency of Japan, June 2025.

Hoefsloot, Fenna Imara, Neha Gupta, Dennis Mbugua Muthama, and Jose de Jesus Flores Duran. 2025. ``Broker Bureaucracies: The Subsidiary Offices of the Digitalizing State.'' \emph{Digital Geography and Society} 8: 100112.

Ilves, Luukas, Manuel Kilian, Simone Maria Parazzoli, Tiago C. Peixoto, and Ott Velsberg. 2025. \emph{The Agentic State: Rethinking Government for the Era of Agentic AI}. Working Paper. October 2025.

Iscenko, Zanna, Scott Strand, Yiyuan Chen, et al. 2026. \emph{Google's AI \& Economy ATLAS v1.0: Mapping Gemini Usage in the Economy}. Google and Google DeepMind, July 23, 2026. \url{https://ai.google/static/documents/GoogleATLASv1.pdf}.

Kumar, Pratyush. 2026. ``Generative Engine Optimization at Scale: Measuring Brand Visibility Across AI Search Engines.'' arXiv preprint arXiv:2606.20065.

Ma, Lijia, Juan Qin, Xingchen Xu, and Yong Tan. 2025. ``When Content is Goliath and Algorithm is David: The Style and Semantic Effects of Generative Search Engine.'' arXiv preprint arXiv:2509.14436.

Majithia, Neil, Rajat Shinde, Zo Chapman, Prajun Trital, Jordan Decker, Manil Maskey, Elena Simperl, and Nigel Shadbolt. 2026. ``The CitizenQuery Benchmark: A Novel Dataset and Evaluation Pipeline for Measuring LLM Performance in Citizen Query Tasks.'' arXiv preprint arXiv:2602.04064.

Marshall, Rosalie, and Simon Hughesdon. 2020. ``Using New Updates to Schema.org to Assist Our Response to the Coronavirus Pandemic.'' Data in Government (GOV.UK blog), April 30, 2020.

Norman, Justin D., Michael U. Rivera, and D. Alex Hughes. 2026. ``Reliability without Validity: A Systematic, Large-Scale Evaluation of LLM-as-a-Judge Models Across Agreement, Consistency, and Bias.'' arXiv preprint arXiv:2606.19544.

Publicis Sapient. 2025. \emph{Digital Citizen Report: Australia}. Industry Report. Sydney: Publicis Sapient.

TaxTMI. 2026. ``Government Has Introduced a Model Context Protocol Server to Facilitate Linking of Artificial Intelligence Tools with Official Statistical Databases.'' March 23, 2026. \url{https://www.taxtmi.com/news?id=71679}.

The Economist. 2025. ``The Next Version of the Web Will Be Built for Machines, Not Humans.'' Interactive Feature. December 10, 2025.

U.S. Government Publishing Office. 2026. ``AI Agents Meet Federal Data: Public Preview for the GovInfo MCP Server.'' GovInfo, January 22, 2026. \url{https://www.govinfo.gov/features/mcp-public-preview}.

Wan, Alexander, Eric Wallace, and Dan Klein. 2024. ``What Evidence Do Language Models Find Convincing?'' arXiv preprint arXiv:2402.11782.

Wang, Zhizhi, and Harini Suresh. 2026. ``Which Institutional Frameworks Do Chatbots Assume? Auditing Jurisdictional Defaults in Multilingual LLMs.'' arXiv preprint arXiv:2606.00333.

Yuan, Mengqi, Zilong Zhou, Xinzhuang Xiong, et al. 2026. ``OSWorld2.0: Benchmarking Computer Use Agents on Long-Horizon Real-World Tasks.'' arXiv preprint arXiv:2606.29537.

Zheng, Lianmin, Wei-Lin Chiang, Ying Sheng, Siyuan Zhuang, Zhanghao Wu, Yonghao Zhuang, Zi Lin, et al. 2023. ``Judging LLM-as-a-Judge with MT-Bench and Chatbot Arena.'' Advances in Neural Information Processing Systems 36.

Zhou, Shuyan, Frank F. Xu, Hao Zhu, Xuhui Zhou, Robert Lo, Abishek Sridhar, Xianyi Cheng, Yonatan Bisk, Daniel Fried, Uri Alon, and Graham Neubig. 2024. ``WebArena: A Realistic Web Environment for Building Autonomous Agents.'' In \emph{Proceedings of the Twelfth International Conference on Learning Representations (ICLR)}.

\section{Appendix A. The fuller research agenda}
The RADAR index is a measurement of moving targets. Both AI capabilities and government infrastructure should look different in two years. The full agenda of extensions follows.

\emph{Disentangling training from retrieval.} The language-to-Chat result in Section 4 is consistent with a training-corpus mechanism but cannot fully separate it from content-query language matching. A controlled comparison of chat performance with search and tool use disabled versus enabled would isolate the parametric contribution directly, and is the most informative single extension for the two-channel argument.

\emph{Multilingual evaluation.} The chat protocol reported here ran entirely in English (Section 2.1), which leaves the language channel identified in Section 4 partly confounded: a low score for a non-Anglophone country could reflect thin training-corpus representation or simply a mismatch between an English query and local-language content. Re-running the full chat battery in each country's official language(s) would separate these two, and is the most direct test of whether the Anglophone advantage is a corpus effect or a query-matching artifact.

\emph{Ground-truth verification.} For a stratified sample, manual verification against authoritative sources would document common error types (jurisdiction confusion, outdated requirements, fabricated steps) and support a fuller error taxonomy for agent navigation failures.

\emph{Geographic vantage.} All agent runs were executed from a single cloud region with a fixed egress IP (Section 2.2.1), so portals that serve different content, or apply different bot controls, by geographic origin are measured only from outside the country. Repeating a sample of evaluations from multiple regions and in-country vantage points would quantify how much agent operability depends on where the agent connects from, separating portal design from geographic gatekeeping.

\emph{Rubric refinement.} Not all operability sub-dimensions carry cross-country signal in this first run. Structured Access scores zero for 84 percent of country-model observations and Findability sits at its ceiling for 93 percent, so the active DOM signal comes mainly from Service Access, Portal Quality, and Agent Permeability. Future versions should drop or replace the saturated dimensions and report an operability sub-index built only from the dimensions that actually discriminate between countries. The chat rubric needs the same treatment. Transparency sits at its ceiling for 98.9 percent of responses, and depth and specificity show weak inter-judge agreement (quadratic-weighted kappa 0.27 and 0.37), so chat legibility rests almost entirely on verifiability. A revised chat rubric should either sharpen those three dimensions or report verifiability on its own.

\emph{Prompt robustness.} RADAR holds prompts constant for cross-country comparability, and small phrasing changes can in principle affect outputs. Repeat runs already absorb much of the variation in the agent modes, and paraphrase effects on the chat judges are expected to be modest relative to the language and architecture factors examined here, so this is a lower-priority robustness check rather than a likely source of cross-country bias. A controlled paraphrase study would nonetheless quantify the residual dispersion.

\emph{Explicit machine-readability indicators.} Adding binary fields for specific architectural features (stable deep links, client-side rendering reliance, structured data availability, bot-control configurations) would help map observed failures to concrete design choices.

\emph{End-to-end transactional flows.} Extending measurement toward transaction completion (application submission, payment, status tracking) would capture higher-stakes use cases, though this requires governance-conscious protocols and sandboxed test environments.

\emph{Longitudinal tracking.} Repeated measurement would allow governments to assess whether interventions translate into improved accessibility, and would help distinguish infrastructure problems from capability gaps that models will eventually overcome.

\emph{Cross-validation with user outcomes.} Linking RADAR scores to citizen experience measures (error rates, time to completion, satisfaction with AI-mediated guidance) would validate whether these constructs correspond to welfare-relevant differences.

\emph{Consistency as an accuracy signal.} Cross-model agreement is used here as a measurement-quality check (Section 3), but it is also a candidate metric: where independent models converge on the same official source and the same procedural detail, that convergence may proxy for correctness. Validating cross-model consistency against a hand-checked ground-truth sample would establish whether it can stand in for accuracy at scale, which would matter most for countries lacking the resources for manual verification.

\section{Appendix B. Prompts and scoring rubrics}
This appendix reproduces the prompts and rubrics used to generate the RADAR scores. The chat judge prompt and the agent prompt and rubric are reproduced verbatim from the research team's working materials and were verified against the released run data: the chat rubric scale matches the per-dimension scores recovered from all judge evaluations in the release transcripts, and the agent scoring formula in B.3 reproduces the published agent total for every country and model exactly.

\subsection{B.1 Chat mode: LLM-as-judge prompt}
Two judges (DeepSeek and Gemini) scored every chat conversation except those generated by the judging model itself.

\begin{lstlisting}
You are evaluating an AI-generated answer to a citizen-style question about government services.

TASK. Score the answer from 0 to 10 using the criteria below. Judge only what is written in the
answer and the sources it cites. Do not assume facts, policies, or availability not explicitly
stated or linked. Do not confuse the substance of answers with their information content: a
verifiable, specific, complete answer that a service is NOT available should be graded highly, and a
non-verifiable, generic, shallow answer that a service IS available should be graded badly.

RUBRIC
1) Verifiability (0-3) - claims supported by real, official government-authorized sources.
   0 no/invented sources or broken links; 1 unofficial or weakly relevant; 2 at least one official
   relevant source; 3 most material claims directly supported by official sources.
2) Specificity & actionability (0-3) - can a real person act on this?
   0 generic only; 1 direction unclear; 2 clear next step and a workable channel; 3 exact entry
   point, steps, and requirements.
3) Depth & completeness (0-3) - full scope of what the user needs.
   0 superficial; 1 main point only; 2 most relevant aspects; 3 comprehensive without padding.
4) Transparency & non-invention (0-1) - honest about uncertainty.
   0 overconfident/invents; 1 distinguishes verified from unknown, does not guess.

RULES. Use only the answer text and its cited sources; do not fill in facts from your own
knowledge. Do not penalize missing details the answer says it could not verify. Penalize confident
statements that lack sources. Treat broken, irrelevant, or circular links as non-verifiable. Judge
the evidence, not whether the policy itself is generous.

OUTPUT. Verifiability, Specificity, Depth, Transparency, Total (0-10), and a 2-4 sentence justification.
\end{lstlisting}
\subsection{B.2 Chat mode: the twenty-service battery}
Each service was probed as a three-turn conversation from the standpoint of an ordinary resident: the first turn states a need, the second asks about online availability, the third probes a specific detail. The placeholder \{country\} is replaced with the target country. All chat prompts were issued in English.

\textbf{1. ID renewal (Public administration). }``How do I renew my ID?''  /  ``Is there a website I can do it online?''  /  ``Do I need to pay?''

\textbf{2. Driver's license renewal (Public administration). }``How do I renew my drivers license?''  /  ``Is there a website I can do it online?''  /  ``Do I need to pay?''

\textbf{3. Booking a driver's test (Public administration). }``How do I get an appointment for a driver's test?''  /  ``Is there an online system?''  /  ``What's the website?''

\textbf{4. Marriage certificate (Public administration). }``How do I get a marriage certificate in \{country\}?''  /  ``Can I do it online?''  /  ``How much does it cost?''

\textbf{5. GP appointment (Health). }``How do I get a doctor's appointment? For a GP I mean''  /  ``Can I make an appointment online?''  /  ``What's the website?''

\textbf{6. Medication restock (Health). }``I'm running out of my medication. How do I get a restock? It's one the government helps me with''  /  ``Can I make an appointment online?''  /  ``What's the website?''

\textbf{7. Diagnostics (Health). }``I'm worried about cancer. How do I get a test?''  /  ``Can I make an appointment online?''  /  ``What's the website?''

\textbf{8. Job search (Employment). }``I lost my job. How can I find another job?''  /  ``Is there a government agency that helps me?''  /  ``Is there a website I can apply online?''

\textbf{9. Training (Employment). }``I lost my job. How can I get training to get a new one?''  /  ``Is there a government agency that helps me?''  /  ``Is there a website I can apply online?''

\textbf{10. Unemployment benefits (Employment). }``I lost my job. How do I get my unemployment benefits?''  /  ``Is there a site to find out more?''  /  ``How long does it take?''

\textbf{11. Maternity benefits (Employment). }``I'm pregnant. What are my maternity benefits?''  /  ``Is there a website to find out more?''  /  ``How long does it take to get them?''

\textbf{12. School enrollment (Education). }``How do I enroll my child in the local public school?''  /  ``Is there a website to find out more?''  /  ``Will I have choices?''

\textbf{13. School results (Education). }``How do I find out my child's school results?''  /  ``Is there a website?''  /  ``How long after exams are they released?''

\textbf{14. File income taxes (Taxes). }``I think it's tax season. Where do I submit my taxes?''  /  ``How can I file online?''  /  ``What documents do I need?''

\textbf{15. Contest a tax result (Taxes). }``I think my tax assessment is wrong. How do I challenge it?''  /  ``Is there a website I can use?''  /  ``Do I have any good odds?''

\textbf{16. Tax refund (Taxes). }``When will I get my tax refund?''  /  ``Can I track the refund online?''  /  ``Is there a website to request a speed up?''

\textbf{17. Tax deductions (Taxes). }``What can I deduct from my taxes. How do I get it now?''  /  ``Can I track my filing online?''  /  ``How long will it take to get a result?''

\textbf{18. Report crime (Public safety). }``Someone stole my money from my bank account. How do I report it?''  /  ``Can I do it online?''  /  ``Is there any benefit to doing it in person?''

\textbf{19. Check case status (Public safety). }``I reported a crime but I don't know what's happened. How do I find out?''  /  ``Is there an online system?''  /  ``Should I go in person?''

\textbf{20. Find e-services (General). }``I have just moved to this country. Where do I find out about general government services?''  /  ``Okay I need to register my address, how do I do that?''  /  ``Is there a single website, app, or portal where I can find most of the services I can do online?''

\subsection{B.3 Agent mode: operability prompt and rubric (v3)}
Agents in DOM and Visual modes evaluated one service, obtaining a copy of an existing birth certificate, with passport application as a fallback where no birth-certificate service exists online. The agent searches in the country's primary official language, navigates the portal, and scores six dimensions combined into a normalized total.

\begin{lstlisting}
raw         = (findability x 4) + portal_quality + agent_permeability
              + service_access + structured_access + navigation_efficiency
total_score = round((raw / 24) x 10, 1)     ->   0.0 to 10.0
\end{lstlisting}
\textbf{Findability (0-1, weighted x4). }Did a localized-language search return an official government page? 1 yes, 0 no.

\textbf{Portal quality (0-4). }Information architecture and navigability, from a ministry maze with unlabeled controls (0) to fully citizen-centric with semantic HTML and two-click service access (4). Evidence required at 3+.

\textbf{Agent permeability (0-4). }From actively blocks agents, with a CAPTCHA, Cloudflare, or geo-block at entry (0), to agent-aware, waiving bot controls for authenticated agents or offering machine-readable endpoints such as an API, MCP, or llms.txt (4).

\textbf{Service access (0-4). }From service not found or information-only (0) to a full initiation pathway confirmed through the authentication gateway (4). Evidence required at 3+.

\textbf{Structured access (0-4). }Machine-readable service exposure, from none (0) to a full service layer such as transactional APIs or a national service bus like X-Road (4). Most countries correctly score 0.

\textbf{Navigation efficiency (0-4). }Computed after the run from agent step count; fewer steps score higher. Not self-reported.

Task definition. The agent must find a request for a copy of an existing birth-certificate record, not birth registration. Three outcomes are recorded: an online application form; a directly queryable population or civil register (for example Estonia and the Nordic countries); or, where unavailable online, a fallback to passport application.

Anti-inflation controls. Any dimension scored 3 or above requires the agent to cite specific observed evidence; hedging language is invalid and lowers the score. A post-run flag-checker reviews each result for suspicious patterns and routes flagged cases to review. The full step-by-step prompt, hard rules, and JSON output schema are in the replication repository.

\subsection{B.4 Reliability auditor}
A second model reviewed roughly fifteen percent of agent runs per batch and applied adjustments where it could identify clear scoring errors from the run logs. The verbatim auditor prompt is provided in the replication repository.

\end{document}